\documentclass[journal]{IEEEtran}

\usepackage{url}
\usepackage{ifthen}
\usepackage[cmex10]{amsmath} 
\usepackage{multirow}
\usepackage{color}
\usepackage{soul}
\usepackage{booktabs}
\usepackage{pbox}
\usepackage{subfigure}
\usepackage[pdftex]{graphicx}
\usepackage{cite}
\usepackage{array}
\usepackage{amsmath}
\usepackage{pifont}
\usepackage{amssymb}
\usepackage{algpseudocode,algorithm,algorithmicx}

\algrenewcommand\algorithmicrequire{\textbf{Initialize:}}
\algrenewcommand\algorithmicensure{\textbf{Postcondition:}}

\def\X{{\cal X}}
\def\I{{\cal I}}
\def\E{{\mathbb E}}
\begin{document}
\title{Constellation Shaping for WDM systems using 256QAM/1024QAM with Probabilistic Optimization}

\author{Metodi P. Yankov,~\IEEEmembership{Member,~IEEE}, 
Francesco Da Ros~\IEEEmembership{Member,~IEEE,~OSA}, 
Edson P. da Silva~\IEEEmembership{Student Member,~IEEE,~OSA}, 
S\o ren Forchhammer~\IEEEmembership{Member,~IEEE}, 
Knud J. Larsen, 
Leif K. Oxenl\o we~\IEEEmembership{Member,~OSA}, 
Michael Galili~\IEEEmembership{Member,~IEEE}, and 
Darko Zibar~\IEEEmembership{Member,~IEEE} 
\thanks{The authors are with the Department
of Photonics Engineering, Technical University of Denmark,
2800 Kgs. Lyngby, Denmark: e-mail: meya@fotonik.dtu.dk}
\thanks{Copyright (c) 2016 IEEE. Personal use of this material is permitted.  However, permission to use this material for any other purposes must be obtained from the IEEE by sending a request to pubs-permissions@ieee.org}
}

\maketitle

\begin{abstract}
In this paper, probabilistic shaping is numerically and experimentally investigated for increasing the transmission reach of wavelength division multiplexed (WDM) optical communication systems employing quadrature amplitude modulation (QAM). An optimized probability mass function (PMF) of the QAM symbols is first found from a modified Blahut-Arimoto algorithm for the optical channel. A turbo coded bit interleaved coded modulation system is then applied, which relies on many-to-one labeling to achieve the desired PMF, thereby achieving shaping gains. Pilot symbols at rate at most 2\% are used for synchronization and equalization, making it possible to receive input constellations as large as 1024QAM. The system is evaluated experimentally on a 10 GBaud, 5 channels WDM setup. The maximum system reach is increased w.r.t. standard 1024QAM by 20\% at input data rate of 4.65 bits/symbol and up to 75\% at 5.46 bits/symbol. It is shown that rate adaptation does not require changing of the modulation format. The performance of the proposed 1024QAM shaped system is validated on all 5 channels of the WDM signal for selected distances and rates. Finally, it is shown via EXIT charts and BER analysis that iterative demapping, while generally beneficial to the system, is not a requirement for achieving the shaping gain.
\end{abstract}


\IEEEpeerreviewmaketitle

\section{Introduction}
High order modulation formats have been a hot topic for investigation for the fiber optic channel community in the last decade due to their capability of transmission with high spectral efficiency (SE). In such cases, uniform input distribution suffers from shaping loss, which for the linear AWGN channels and high SNR is found to be $1.53$dB \cite{Cover}. The fiber optic channel on the other hand is highly non-linear in the high signal-to-noise ratio (SNR) regime, which limits the maximum achievable SE \cite{Essiambre} with standard receivers and makes the ultimate shaping gain analysis difficult. Particularly so due to the memory effect of chromatic dispersion (CD) and its interaction with the non-linear self and cross phase modulation (SPM and XPM, respectively) in the fiber, which are not considered in standard approaches for estimating and maximizing the mutual information (MI) between channel input and output.  

In \cite{Dar-shaping}, the memory is taken into account, and the probability density function (PDF) of a continuous input is optimized to a multidimensional ball. This leads to potential shaping gains exceeding the above-mentioned $1.53$dB, suggesting similar results for discrete constellations of large size, e.g. high order quadrature amplitude modulation (QAM). However, memory-aware receivers generally require exponential increase in complexity. If a typical value of 16 symbols is used to express the channel memory (see e.g. \cite{Dar-shaping}) in combination with a constellation with 16 points, the receiver would generally have to go through all $2^{4\cdot16}$ possible states of the input, something clearly impractical. 
The independent, identical distribution (i.i.d.) assumption on the input symbols, even though underestimating the channel capacity, provides a proper estimate of the achievable rates by current digital systems, which usually do not consider the memory. Under this assumption, well known digital communication techniques can be used to optimize the transmission strategy, among these probabilistic shaping. 

In \cite{Smith}, trellis shaping is used to achieve very high SE, close to the theoretical lower bound. Probabilistic shaping is performed in \cite{Fehen} in combination with a low-density parity check (LDPC) code, where the reach of the link is increased by between two and four spans for a long haul link. In \cite{Smith} and \cite{Fehen}, digital back-propagation is performed, a technique which will often be impractical, especially in wavelength division multiplexed (WDM) network scenarios where the receiver does not have access to the interfering channels. In \cite{Agrell}, shell-mapping is used to achieve shaping gain, which suffers from a poor complexity scaling with the constellation size. Perfect synchronization and equalization is assumed in \cite{Smith, Fehen, Agrell}. The system from \cite{Fehen} is later experimentally demonstrated \cite{Buchali, Buchali-2} and significant reach increase is shown in the presence of QPSK interfering signals. Rate adaptation with that scheme is realized by changing the input distribution, which may not always be desirable in WDM networks.

Methods also exist for geometric shaping of the input PDF \cite{Lotz, Estaran, Liu, Djordjevic}. For example in \cite{Lotz}, polar modulation is used, which achieves significant shaping gain over uniformly distributed QAM constellations. Superposition coded modulation was used in \cite{Estaran} as a shaping method. Geometric shaping schemes generally require high-precision digital-to-analog (DAC) and analog-to-digital conversion (ADC) due to the non-uniformity of the signal on the complex plane. Standard QAM constellations are desirable due to their simplicity and the possibility to directly apply I/Q modulation/demodulation.  
  
As part of our previous work \cite{Yankov}, a near-capacity achieving coded modulation scheme was designed for an AWGN channel, based on turbo-coded bit-interleaved coded modulation (BICM) employing QAM and constellation shaping. In \cite{Yankov-PTL}, the QAM probability mass function (PMF) is optimized for a single channel optical system and up to 1.5 dB gains are numerically shown in both the linear and non-linear region of transmission. 

In this paper, the single channel work from \cite{Yankov-PTL} is extended to WDM scenarios, where both intra- and inter-channel non-linearities are present. We also perform a distance study, where the shaping gain is evaluated in terms of transmission reach. We evaluate the achievable MI rate after the synchronization and equalization, and show that the proposed turbo-coded scheme closely approaches it. 

The main contribution of this paper is the experimental demonstration of the method in a WDM optical fiber system with 5 channels. The performance on all channels is validated for selected scenarios.  

\section{Theoretical background}
\label{section:theory}
In information theory, channel capacity is defined as the maximum MI between the channel input and output, where the maximization is performed over all input PDF (PMF in the discrete case). For channels with memory the capacity can be found as \cite[Eq.(41)]{Essiambre}
\begin{align}
 C &= \lim_{K \rightarrow \infty} \max_{p_{X_1^K}(X_1^K)} \frac{1}{K} {\cal{I}}(X_1^K;Y_1^K)  \notag \\
 \label{eq:capacity}
 &= \lim_{K \rightarrow \infty} \max_{p_{X_1^K}(X_1^K)} \left[ \frac{1}{K} {\cal{H}}(X_1^K) - \frac{1}{K} {\cal{H}}(X_1^K|Y_1^K) \right] ,
\end{align}
where $X_1^K$ and $Y_1^K$ are the channel input and output sequences of length $K$, respectively, ${\cal{I}}$ is the MI, and ${\cal{H}}$ is the entropy function. The output sequence of the channel can be obtained by solving the non-linear Shr\"odinger equation (NLSE) by the split-step Fourier method (SSFM). The input-output relation is governed by some probability function $p_{X_1^K|Y_1^K}(X_1^K|Y_1^K)$, which for the fiber is generally unknown. If the conditional entropy in Eq.~(\ref{eq:capacity}) is defined to include the true underlying PMF $p$ as ${\cal{H}}_p(X_1^K|Y_1^K)$, and instead, an auxiliary PMF $q_{X_1^K|Y_1^K}(X_1^K|Y_1^K)$ is used for the entropy calculation, the result is an upper bound: $\bar{ {\cal{H}}}(X_1^K|Y_1^K) = {\cal{H}}_q(X_1^K|Y_1^K) \ge {\cal{H}}_p(X_1^K|Y_1^K)$, with equality iff $q_{X_1^K|Y_1^K}(X_1^K|Y_1^K) = p_{X_1^K|Y_1^K}(X_1^K|Y_1^K)$ \cite{Arnold}. In particular, disregarding the memory effect in the channel corresponds to an auxiliary function $q_{X_1^K|Y_1^K}(X_1^K|Y_1^K) = \prod_{k=1}^K p_{X|Y}(X_k|Y_k)$, where $X_k$ and $Y_k$ represent the input and output samples of the channel at time $k$, respectively.  When constraining the input to i.i.d. symbols and using this auxiliary function, a lower bound on the capacity is found as
\begin{gather}
\label{eq:MI_LB}
 C \ge \hat{C} = \max_{p_X(X)} \left[ {\cal{H}}(X) - \bar{{\cal{H}}}(X|Y) \right] ,
\end{gather}
where $X$ and $Y$ are scalars, $X$ comes from the constellation $\cal{X}$, which in our case will be QAM, and $Y \in {\mathbb{C}}$. The PDFs in the remaining of the paper are assumed memoryless.

A typical receiver will attempt to find a symbol sequence $\hat{x}_1^K$, which is optimal in some sense. For example, the maximum likelihood solution is given by $\hat{x}_k = \arg \max_{X} q_{Y|X}(Y|X)$, whereas the maximum a-posteriori probability solution is $\hat{x}_k = \arg \max_{X} q_{X|Y}(X|Y) = \arg \max_{X} p_X(X)  q_{Y|X}(Y|X)$. In either case, some sort of auxiliary channel function $q_{Y|X}$ is employed for demodulation and decoding, the quality of which governs the performance of the receiver. The lower bound in (\ref{eq:MI_LB}) will thus represent the maximum error-free data rate which can be achieved by the specific receiver, employing a specific auxiliary function. The lower bounds is therefore also referred to as achievable information rate (AIR).  

\subsection{Probabilistic optimization}
\label{sec:optimization}
The standard approach to perform the optimization in Eq.~(\ref{eq:MI_LB}) for the AWGN channel with constellation constrained input is the Blahut-Arimoto algorithm \cite{Varnica}. The algorithm is straight-forward in that case due to the simple likelihood function, which is the Gaussian PDF with the variance of the additive noise. Since the input-output relation of the fiber-optic channel is not available in closed form, and is furthermore signal dependent, we model the memoryless auxiliary PDF $q_{Y|X}(Y|X=x^i)$ as a 2D Gaussian distribution ${\cal N}(\Sigma_i, \mu_i; \left[ \mathrm{Re} \left[Y\right] , \mathrm{Im} \left[Y\right] \right]^T)$ with covariance matrix $\Sigma_i$ and mean $\mu_i$, where $x^i$ is the $i-$th element in $\X$ \cite{Eriksson}. The parameters of the Gaussians are estimated from training sequences $\{ x_1^K, y_1^K \}$ by taking the sample mean and covariance matrix of the sub-set of points $y_{{\cal K}_i}$, where ${\cal K}_i$ is the set of indexes $k$ for which $x_k = x^i$. The modified Blahut-Arimoto algorithm for finding an optimized input PMF for the optical channel at each input power $P_{in}$ is given in Algorithm~\ref{alg:optical_BAA}. 
\begin{algorithm}[!b]
\caption{Algorithm for optimizing the PMF
   \label{alg:optical_BAA}}
 \begin{algorithmic}[1]
 \For{$\alpha \in \left[1/\max(|X|); 1/\min(|X|)\right]$6\cite{Varnica}}
   \Require $p_X(X)$ s.t. $\alpha^2\E\left[X^HX\right] = P_{in}$
	\While{ Not converged $p_X(X)$}
	\State Generate $x_1^K \sim p_X(X), X\in \alpha \X$
	\State Generate $y_1^K$ by solving the NLSE via the SSFM
	\For{each $x^i \in {\cal X}$}
	\State Find the set $k \in {\cal K}$ for which $x_k = x^i$. 
	\State Estimate $\Sigma_i = {\mathbb{C}\mathrm{ov}} \left[ {\mathrm{Re}}\left[Y_{{\cal K}}\right], {\mathrm{Im}}\left[Y_{{\cal K}}\right] \right]^T $
	\State Estimate $\mu_i = {\mathbb{E}}_k \left[ {\mathrm{Re}}\left[Y_{{\cal K}}\right], {\mathrm{Im}}\left[Y_{{\cal K}}\right] \right]^T $
	\EndFor
	\State Set $q_{X|Y}(X|Y) = \frac{p_X(X)q_{Y|X}(Y|X)}{\sum_{x^i \in \alpha \X} p_X(X=x^i)q_{Y|X}(Y|X=x^i)}$
	\State $p_X(X) = \arg \max_{p_X(X)} {\cal{I}}(Y;X)$, s.t. $\alpha^2\E\left[X^HX\right] = P_{in}$ and $p_{X|Y}(X|Y) = q_{X|Y}(X|Y)$.
	\EndWhile
    \State $\I^{\alpha} = \I_q(X;Y)$ with $p_X(X)$ and $q_{X|Y}(X|Y)$
    \EndFor
    \State $\hat{C} = \max_\alpha \I^\alpha$
 \end{algorithmic}
\end{algorithm}
The outer-most loop over $\alpha$ in Algorithm~\ref{alg:optical_BAA} allows for a basic geometric shaping via scaling of the original constellation \cite{Varnica}. We observed that less than 5 iterations are required for the PMF to converge, which with $10^5$ symbols results in a few hours runtime for each parameter set $\{ \alpha, P_{in} \}$. We note that accurate MI estimation for constellations larger than 1024QAM would require longer sequences, thereby increasing the runtime of the algorithm significantly. Approximate models, e.g. \cite{Dar_invited, Carena_EGN}, would be of interest for performing the optimization in that case, particularly for replacing Steps 3-10 in Algorithm \ref{alg:optical_BAA}.

\subsection{Dyadic PMF approximation}
\label{sec:dyadic_PMFs}
The optimal PMF as found by Algorithm \ref{alg:optical_BAA} is not trivial to realize when communicating binary data. The authors in \cite{Bocherer-2, Fehen, Buchali, Buchali-2} propose a rate matcher, which for long block lengths is able to approach the desired PMF (in their case that is a Maxwell-Boltzmann (MB) distribution). Instead, here the concept of \textit{many-to-one} mappings \cite{Raphaeli} is adopted, which allows for simple realization of distributions of the form $p_X(X=x_i) = 2^{-l_i}$, where $l_i$ is an integer. Such distributions are referred to as \textit{dyadic}. The dyadic approximation with smallest Kullback-Leibler divergence from the optimal PMF can be found by geometric Huffman coding (GHC) \cite{Bocherer}. In order to ensure a binary-reflected Gray code is possible (see  Section~\ref{sec:many-to-one} and \cite{Yankov}), dyadic PMFs are first found for the marginal PMFs in each I and Q, and the dyadic QAM PMF is constructed by taking their product.

\begin{figure}[!t]
\centering
\includegraphics[width=3.4in]{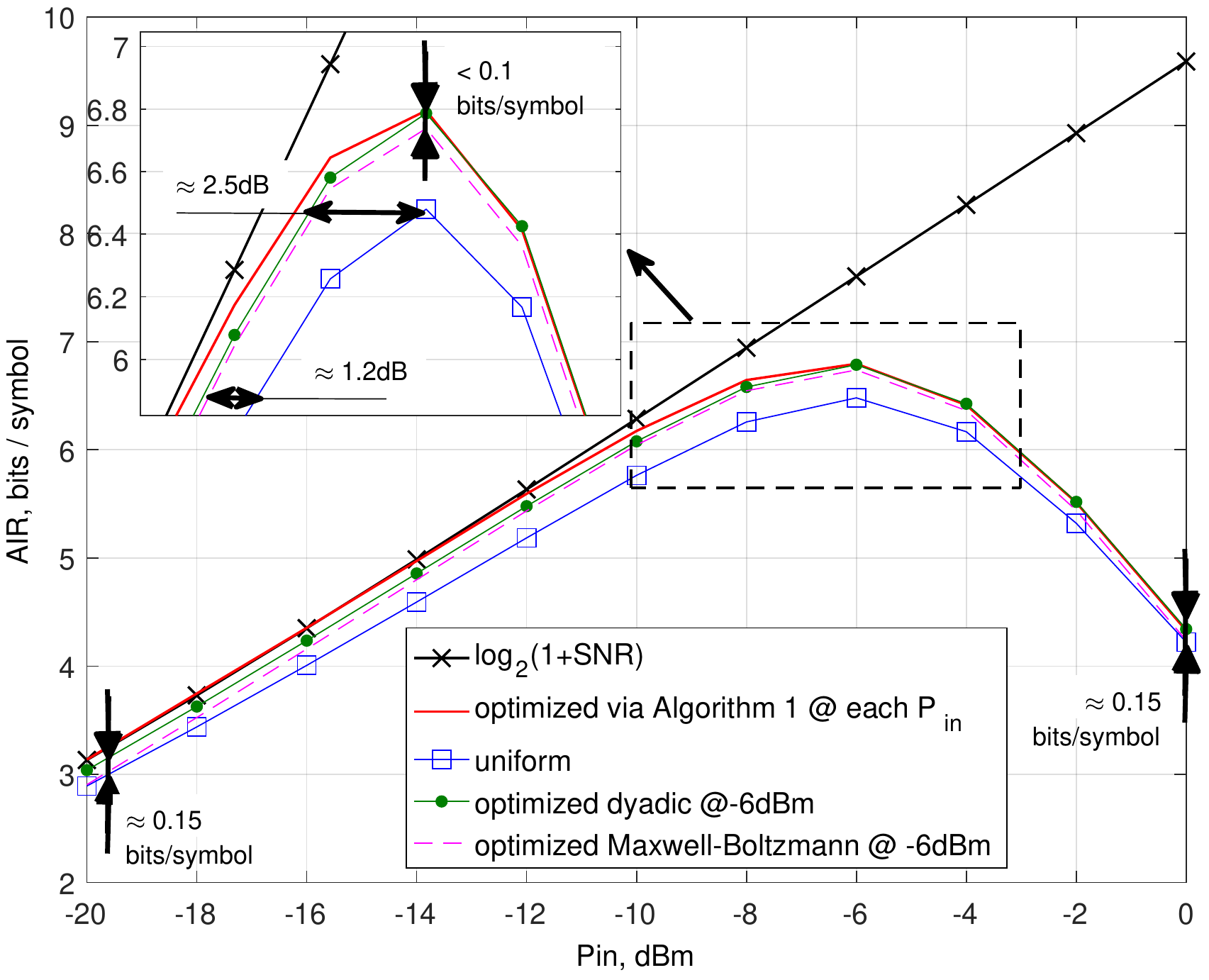}
\vspace{-.2cm}
\caption{Achievable information rates for 1024QAM with the PMF, as obtained from Algorithm \ref{alg:optical_BAA}, uniform PMF, the PMF, which is obtained as the GHC dyadic approximations to the optimal PMF at input power $-6$ dBm, and the Maxwell-Boltzmann PMF, which is optimized at input power $-6$ dBm.} 
\label{fig:rates_basic}
\vspace{-.2cm}
\end{figure}

The PMFs, obtained by the procedure above are examined for an erbium doped fiber amplifier (EDFA) lumped amplification link with parameters, given in Table~\ref{tbl:sim_params}. In Fig.~\ref{fig:rates_basic}, example AIRs for 1024QAM with optimized, dyadic, uniform and MB PMFs are given. The latter is obtained as $p_X(X) \propto \exp(-\lambda |X|^2)$. The performance is given in bits/symbol/polarization (simply bits/symbol from now on). For Nyquist spacing of the WDM channels this corresponds to bits/s/Hz/polarization, or the spectral efficiency of the modulation format. The input power is specified in dBm per channel. The input/output sequences are extracted from one of the polarizations of the central channel. While some rate loss can be expected, i.i.d. polarizations are assumed due to the increased complexity of joint processing \cite{Eriksson}, and the AIRs are calculated similar to $\I_q$ in Algorithm 1 using Steps 3-13.

The PMF is optimized at each input power, while the dyadic PMF is obtained as the GHC approximation to the optimized PMF at input power $-6$ dBm and evaluated on the rest of the region. The MB PMF is similarly optimized by a brute-force search of the parameter $\lambda$ at input power $-6$dBm and evaluated on the remaining region. We see that at that launch power, around 0.4 bits per symbol can be gained by the PMF, optimized with Algorithm 1, but also with its dyadic approximation. This is due to the large constellation size (1024QAM). Smaller constellations suffer from a more significant rate loss from the dyadic approximation. 

In Fig.~\ref{fig:rates_basic}, we also see that the AIR gain w.r.t. the MB PMF is marginal ($< 0.1$ bits/symbol). This indicates that a MB PMF with properly optimized $\lambda$, which is a near-optimal procedure for linear channels, is practically sufficient for application to a (fairly) static nonlinear fiber channel with the constraints, considered here (such as memoryless processing, independent polarizations and channels).

We also observe that in the severely nonlinear and low SNR regions (where the AIR is significantly different) the MB PMF loses its shaping gain and needs to be re-optimized for the corresponding rate. This is not the case for the dyadic PMF, which achieves shaping gains on the entire region ($\approx0.15$ bits/symbols in the extremities).

At the maximum AIR of the uniform PMF, the dyadic PMF achieves around 2.5 dB energy efficiency gains due to the saturation of the AIR. Operating in such points is preferable and indeed possible, as will also be demonstrated. 

\subsection{Many-to-one mapping function}
\label{sec:many-to-one}
\begin{figure}[!t]
\centering
\includegraphics[width=1.4in, angle =90 ]{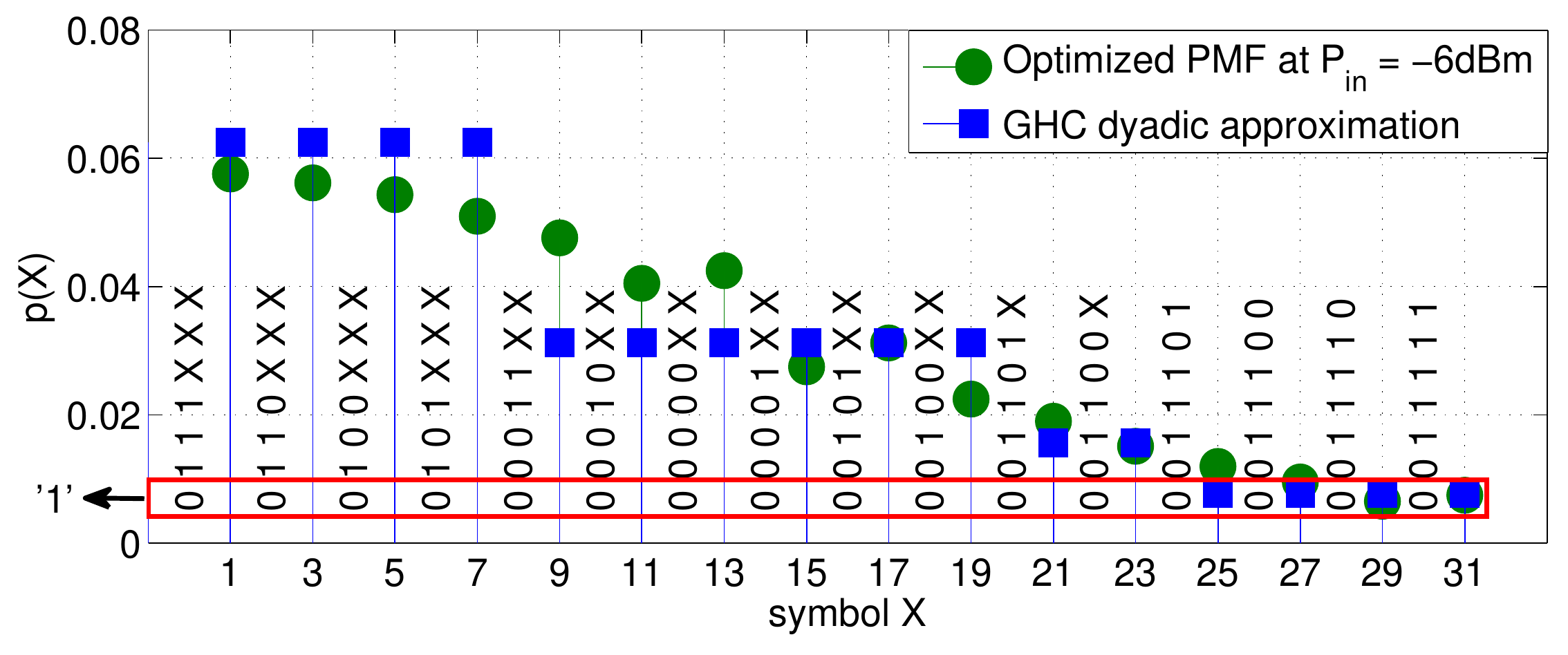}
\vspace{-.3cm}
\caption{Optimized 32PAM PMF, together with the corresponding mapping functions and the dyadic approximation. Only right-hand side. Explanation in the text.}
\label{fig:1024QAM_PMF}
\vspace{-.3cm}
\end{figure}

\begin{figure}[!t]
\centering
\includegraphics[width=1.4in, angle =90 ]{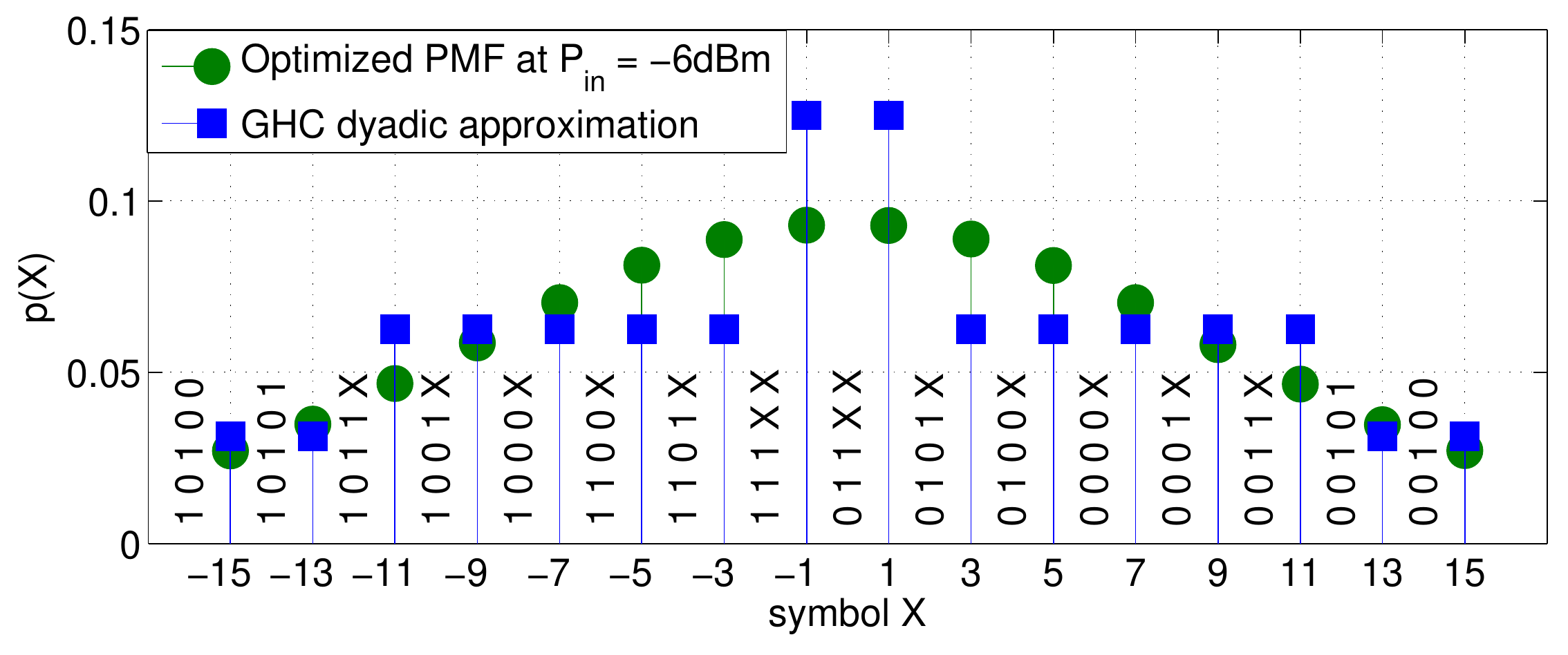}
\vspace{-.3cm}
\caption{Optimized 16PAM PMF, together with the corresponding mapping functions and the dyadic approximation.}
\label{fig:256QAM_PMF}
\vspace{-.3cm}
\end{figure}

Once having the optimal dyadic probabilities of each symbol, the algorithm from \cite{Yankov} can be used to find the desired many-to-one labeling. A many-to-one mapper requires that multiple bit labels are assigned to the same symbol from the constellation. This allows for the bit label length $m>\log_2|{\cal X}|$, as opposed to standard Gray-label QAM, for which $m=\log_2|{\cal X}|$ (e.g. $m=8$ for standard uniform 256QAM). This non-bijective mapping creates ambiguities in some bit positions (to be explained in the following). The proposed labeling for the optimized 1024QAM and 256QAM PMFs are given in Fig.~\ref{fig:1024QAM_PMF} and Fig.~\ref{fig:256QAM_PMF}, respectively. The bits, labeled as 'X' take on both '1' and '0', and the decoder (if successful) will correct for this ambiguity. We only plot the marginal PMF in 1D, i.e. 32PAM and 16PAM, respectively. 

\begin{table}[!t]
\renewcommand{\arraystretch}{1.2}
\renewcommand{\tabcolsep}{15pt}
\caption{System parameters for the optimization}
\label{tbl:sim_params}
\centering
\begin{tabular}{rl}
\hline
\hline
Fiber loss & $\alpha = 0.2$ dB/km \\
Non-linear coefficient & $\gamma = 1.3$ (W$\cdot$km)$^{-1}$ \\
Dispersion & $D = 17$ ps/(nm$\cdot$km) \\
Central wavelenght & $\lambda_0 = 1.55 \text{ } \mu$m \\
SSFM step & $0.1$ km \\
Fiber length & $800$ km \\
Span length & $80$ km \\
EDFA noise figure & $3$ dB \\
Number of channels & $5$ \\
Symbol rate & $28$ GBaud \\
Channel spacing & $30$ GHz \\
Pulse shape & Square root raised cosine \\
Roll-off factor & 0.01 \\
\hline
\hline
\end{tabular}
\end{table}

Due to space considerations and the fact, that the PMFs are symmetric around 0, only half of the mapping function is given for the optimal 1024QAM, i.e. the right side. The left side of the mapping function is obtained by mirroring the right side, while flipping the most significant bit from \lq{}0\rq{} to \lq{}1\rq{} \cite{Yankov}. This symmetry can be seen in Fig.~\ref{fig:256QAM_PMF}. We note the non-Gaussianity of the PMFs, particularly the one from Fig.~\ref{fig:1024QAM_PMF}, which allows for the PMF to perform near-optimally on a wide input power region.

Each symbol can take multiple bit labels, making some symbols more likely to be produced when the input binary data are i.i.d. For example symbol '3' in Fig.~\ref{fig:1024QAM_PMF} takes all bit strings of length 7, which begin with the prefix '0110' (8 different bit strings in total), making its probability $2^{-4}$. Symbol '31' only takes one bit string, and its probability is therefore $2^{-7}$.  

\section{Transceiver details}

\begin{figure*}[!t]
\centering
\includegraphics[width=5.5in]{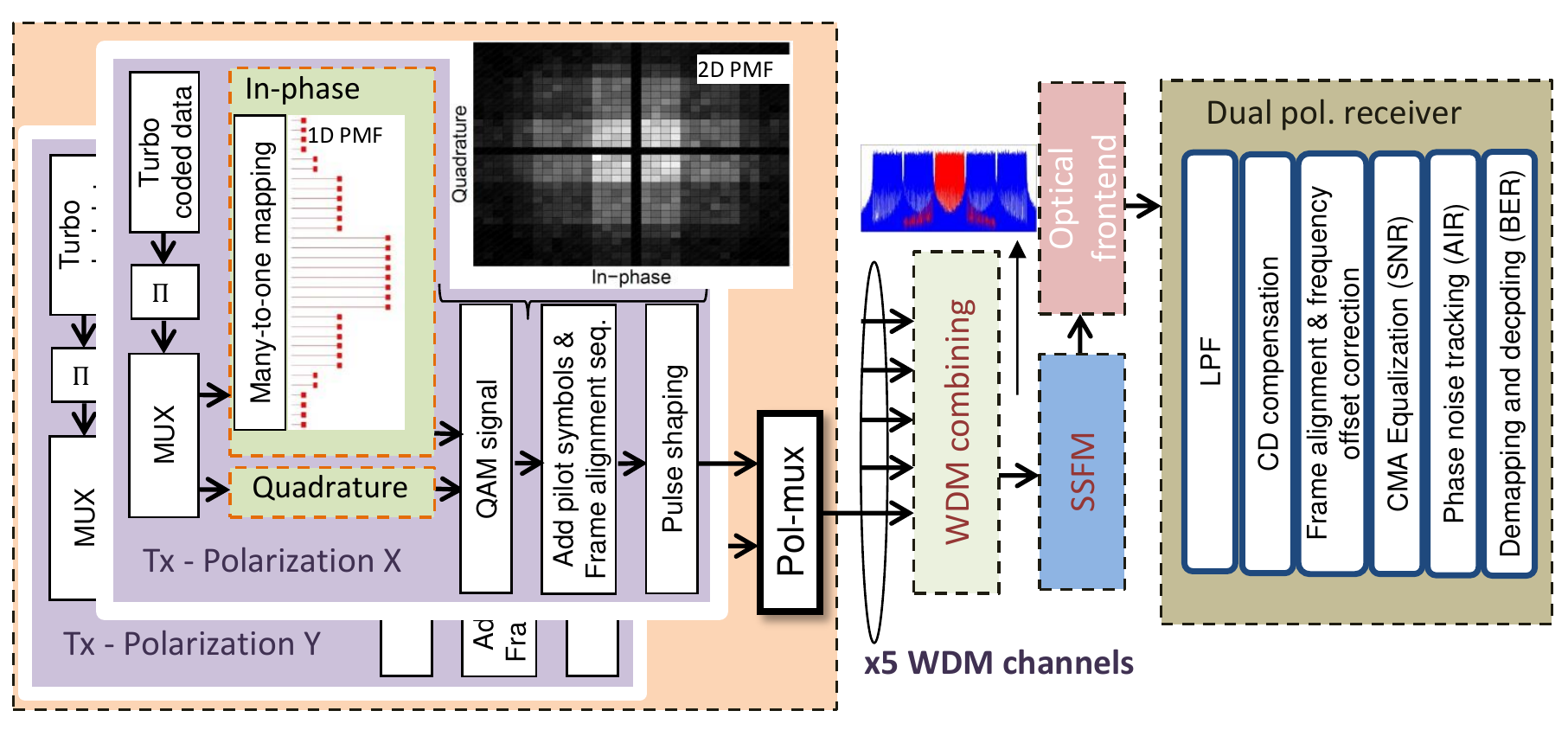}
\vspace{-.4cm}
\caption{Transmitter and receiver for the designed system. At the receiver, the SNR, AIR and BER are estimated after equalization, phase noise mitigation and decoding, respectively.}
\label{fig:transceiver}
\vspace{-.4cm}
\end{figure*}

Block diagrams of the transmitter and receiver are given in the system setup in Fig.~\ref{fig:transceiver}. The binary data are encoded by a turbo code with convolutional constituent encoders with generator polynomials (23, 37) in standard octal notation. A random permutation is used for interleaving between the convolutional encoders, as well as between the turbo encoder and mapper. The detailed coded modulation and mapping/demapping block diagrams may be seen in \cite{Yankov}, and are omitted here. The input data rate is controlled by properly selecting the forward error correction (FEC) overhead for each constellation. If the desired input data rate per QAM symbol is $\eta$, the FEC overhead must be $\frac{(m-\eta)}{\eta} \cdot 100 \%$. Variable overhead is straight-forward to realize with turbo codes by \textit{puncturing}. That is, only part of the overhead is transmitted, and the rest is disregarded, which may be realized with virtually no additional hardware complexity, since a single base circuitry is used for every rate. This is in contrast to standard schemes where a different encoding/decoding circuit is needed for each rate or e.g. \cite{Buchali, Buchali-2}, where a rate matcher is required. In our design, the turbo code has a base code rate of $R=1/3$, or a maximum of 200\% overhead can be transmitted. In the case of e.g. $\eta=5$ bits/symbol and a standard 1024QAM with bit label length of $m=\log_2 1024 = 10$, a 100\% overhead is transmitted. For the 1024QAM labeling from Fig.~\ref{fig:1024QAM_PMF}, the bit label length is $m=14$ and the transmitted overhead is 180\%. In order to ensure best performance in combination with the turbo code, the labeling needs to have unique bits connected to the systematic bits of the code, and at the same time those bits should fulfill a Gray-like criterion \cite{Yankov}. 

An example histogram of the 1024QAM symbols is shown in the transmitter part of Fig.~\ref{fig:transceiver} for the 32PAM  mapping function from Fig.~\ref{fig:1024QAM_PMF}, where bright color indicates high probability and dark color indicates low probability of occurrence. 

Constellations as large as 256QAM and 1024QAM require near-optimal synchronization and equalization. Non-data aided synchronization usually performs poorly for multi-amplitude constellations. To that end, pilot symbols, consisting of random, uniformly spaced QPSK symbols are interleaved with the data symbols. Further, a short Zadoff-Chu sequence \cite{Zepernick} is inserted before the first block for frame alignment, whose overhead is neglected. The symbol sequence with pilots included is pulse-shaped and sent for transmission after combining with the orthogonal polarization and the other WDM channels.

At the receiver, the optical signal is passed through a low-pass filter, chromatic dispersion (CD) compensation is performed in the frequency domain \cite{CDcomp}, and the signal is down-sampled to two samples per symbol. Frame alignment is then performed and the pilot symbols are extracted from the received sequence. Maximum likelihood (ML) frequency offset estimation and synchronization is performed based on the pilots only. The well-known constant modulus algorithm (CMA) is used for equalization of the extracted QPSK pilots. The equalizer taps are then linearly interpolated and applied on the entire sequence. Using training data, the received SNR is estimated after the equalization process.

We model the combined effect of laser and non-linear phase noise as a Wiener process and track it with the Tikhonov distribution based algorithm from \cite{Yankov-PN}. The variance of the Wiener process is also obtained from training data, together with the mean and variance of the likelihood of each symbol from the transmit alphabet, as described in Section \ref{sec:optimization}. The algorithm from \cite{Yankov-PN} directly produces the posterior distributions $q(X|\hat{Y})$, which are needed by the demapper, and also calculates the entropy $\bar{{\cal{H}}}(X|\hat{Y})$ from Eq.~(\ref{eq:MI_LB}), where $\hat{Y}$ are the symbols, sent for de-mapping and decoding.  

Iterative demapping and decoding can be performed in order to improve the performance overall, however, it is not a requirement for the system. Furthermore, the shaping gains are unaffected by the presence of iterative demapping, as we shall see in the results sections. Unless otherwise stated, 5 demapping iterations are performed. The turbo decoder is fixed at 10 internal iterations.

\section{Simulation results}
In this section, we study the AIRs on the link from Table~\ref{tbl:sim_params}. In order to make a proper assessment, we model the main imperfections across the link and receiver, as given in Fig.~\ref{fig:transceiver}. We induce a frequency offset of 50 MHz and model a laser with a linewidth of 10 kHz. Further, uniform ADC quantization is performed with accuracy of 6 bits per real dimension. The received signal is sampled with 80 GHz sampling rate, or 2.85 samples per symbol. The pilot rate is 1\%, which is sufficient for mitigating the induced phase noise. Lasers with broader linewidth would require increased pilot rate. We evaluate the performance on the central WDM channel.

\begin{figure}[!t]
\centering
\includegraphics[width=3.4in]{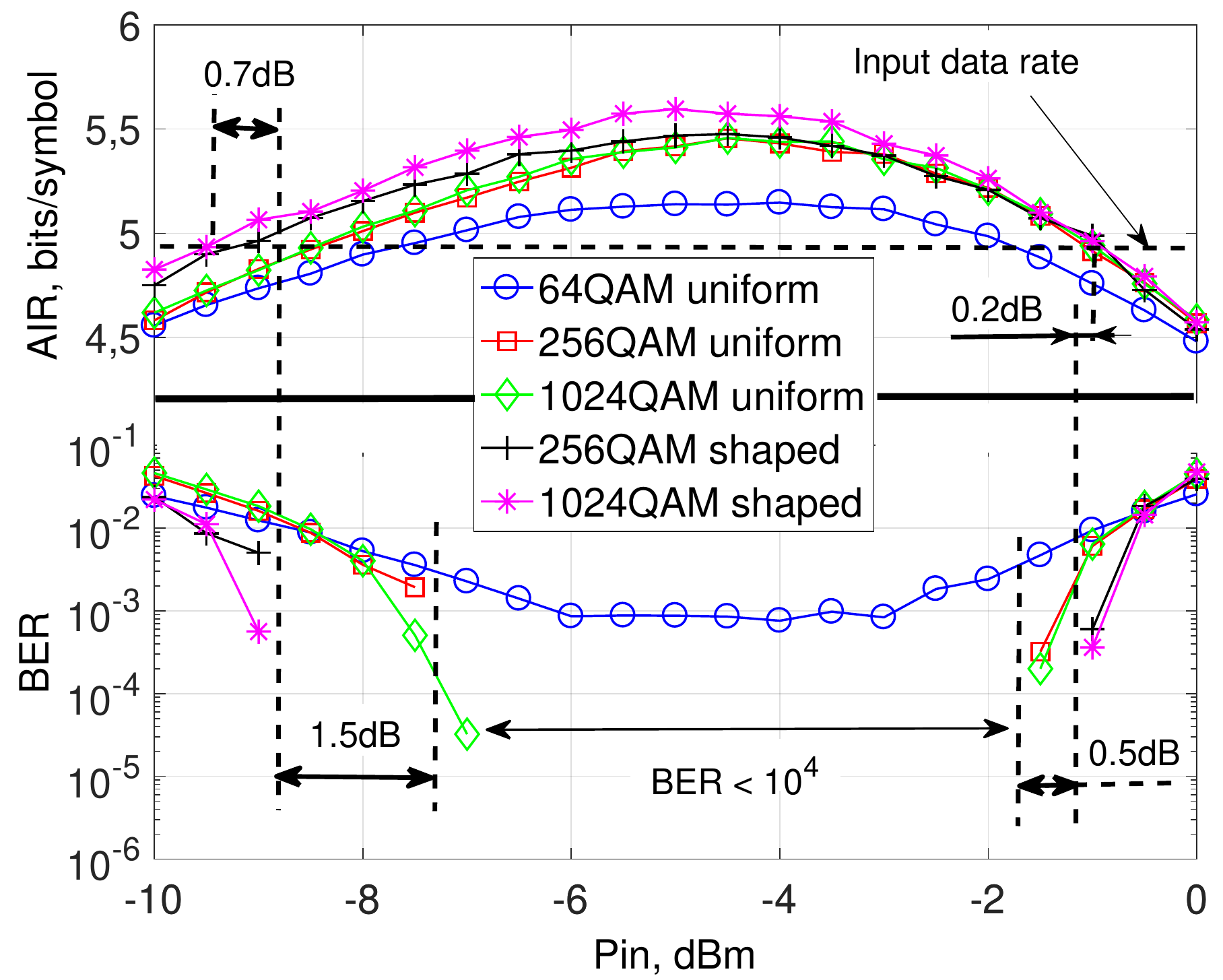}
\vspace{-.2cm}
\caption{\textbf{Simulations:} AIRs and BER for the studied modulation formats at 5 bits/symbol, 800 km. The 64QAM floors at BER$\approx 10^{-3}$. Shaping gain is achieved with both 1024QAM and 256QAM in the linear and non-linear region.}
\label{fig:sim_BER_AIR}
\vspace{-.3cm}
\end{figure}

We study the performance of the 1024QAM and 256QAM labellings from Fig.~\ref{fig:1024QAM_PMF} and Fig.~\ref{fig:256QAM_PMF} and standard Gray labeled uniform 64QAM, 256QAM and 1024QAM. The simulated input data rates are $\eta=5, 5.5, 6 \mbox{ and } 6.5$ bits/symbol. A total of 18 blocks are transmitted in each polarization, each block of length 6000 symbols, making the total number of transmitted bits $\approx 1.08 \cdot 10^6, 1.19 \cdot 10^6, 1.29 \cdot 10^6 \text{ and } 1.40 \cdot 10^6$ for the four rates, respectively. We can therefore safely assume that under stable system operation with stationary noise distribution, the calculated BER after turbo decoding above $10^{-4}$ is reliable, since at least 100 errors are found. Bringing the BER reliably below $10^{-15}$ can be done by adding a 7\% overhead hard decision (HD) FEC code. In Fig.~\ref{fig:sim_BER_AIR}, the BER is shown after 800 km and input data rate $\eta=5$ bits/symbol, together with the AIRs after phase noise tracking as a function of the input power. The constellations from Fig.~\ref{fig:1024QAM_PMF} and Fig.~\ref{fig:256QAM_PMF} achieve around 1.5 dB shaping gain in the linear regime w.r.t. their corresponding uniform constellations, but also around 0.5 dB in the non-linear region. Due to the sub-optimality of the turbo code at this high code rate ($R=5/6$), the 64QAM achieves a minimum BER of $\approx 10^{-3}$. We also see that the turbo code operates at around 0.7 dB and 0.2 dB penalty to the AIRs of the shaped systems in the linear and nonlinear region, respectively, at BER$<10^{-4}$.

\begin{figure*}[!t]
\centering
\includegraphics[width=6.0in]{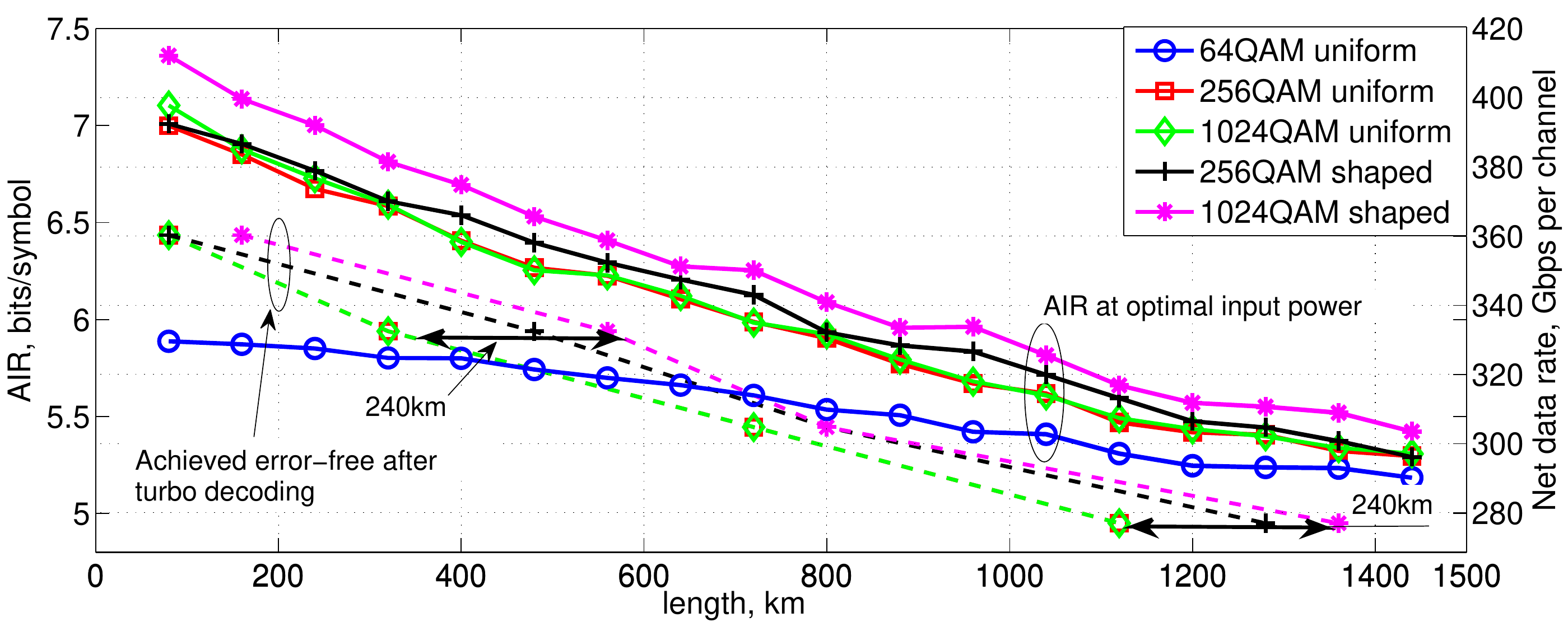}
\vspace{-.4cm}
\caption{\textbf{Simulations:} Maximum AIRs for the studied systems (solid lines) and error-free operating points (dashed lines) after turbo decoding (BER$< 10^{-4}$). The error-free performance is close to AIRs (around 0.4 bits/symbol). The achieved throughput is shown on the right axis.}
\label{fig:sim_rates}
\vspace{-.4cm}
\end{figure*}

In Fig.~\ref{fig:sim_rates}, the AIR and the corresponding throughput after phase noise tracking for each constellation is shown at the respective optimal input power, as a function of the link distance. We note that these AIRs are \textit{independent} of the channel code, and are only a function of the equalization we perform and the input PMF. We see that 64QAM suffers from the limited number of points, and is therefore inferior to the other constellations. The best performing overall constellation is the dyadic 1024QAM from Fig.~\ref{fig:1024QAM_PMF} with distance increase of about 200 km for fixed data rate w.r.t. standard 1024QAM. The 256QAM is also able to achieve shaping gains of around 100 km.  

In Fig.~\ref{fig:sim_rates}, we also show the maximum distances, where error-free transmission is achieved at the different rates (after turbo decoding). We see that shaping provides a steady increase of between 0.3 and 0.5 bits/symbol, or between 8.4 and 14 Gbps per channel. This translates to between 80 km (short distance) and 240 km (long distance) of maximum reach increase at the same data rate. We can also see that the chosen turbo code is able to operate at around 0.6 bits/symbol gap to the maximum AIR.

Both shaped PMFs achieve shaping gain for all studied distances and rates, in contrast to the system from \cite{Buchali}, where the PMF needs to be changed with the distance and/or rate. The system proposed here relies on optimizing the PMF at the \textit{optimal} input power and performing the rate adaptation via puncturing, i.e. the PMF is fixed. This in turn results in constant XPM induced interference in neighboring channels \cite{Dar_invited}, which may be beneficial in highly dynamic network scenarios.

\section{Experimental setup}
\begin{figure*}[!t]
\centering
\includegraphics[width=6.8in]{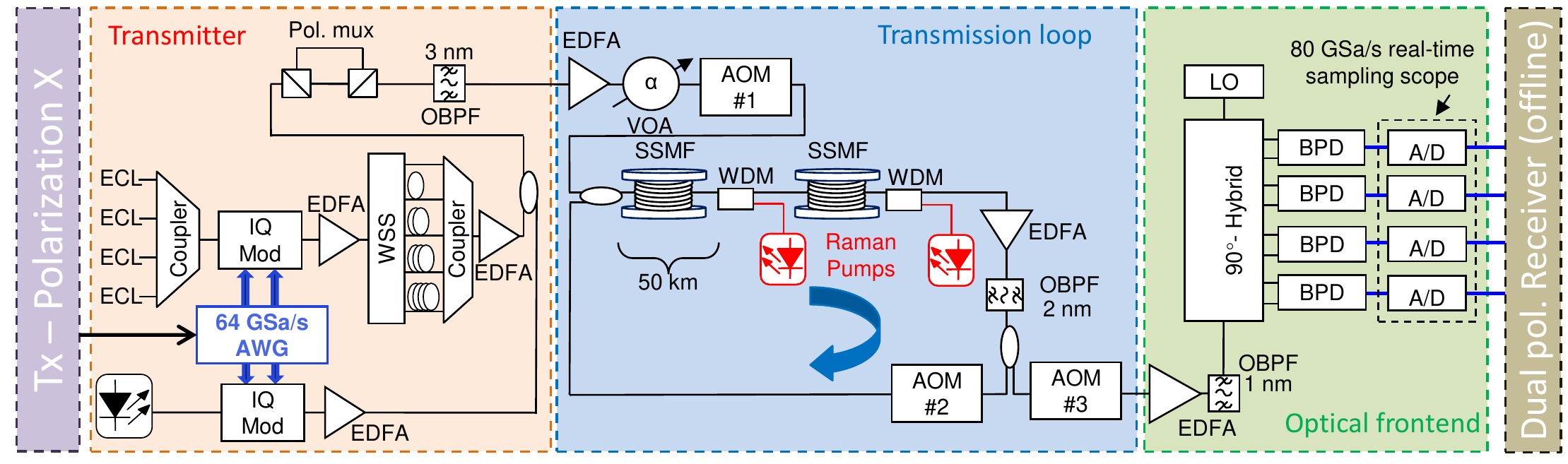}
\vspace{-.2cm}
\caption{Experimental setup. The waveform of one polarization is generated offline, then fed to the AWG. WDM signal is then generated, and sent to the 100 km DRA recirculating loop. After 80 GSa/s coherent reception, the received samples are processed offline.}
\label{fig:exp_setup}
\vspace{-.2cm}
\end{figure*}

The proposed system is next experimentally validated. To that end, the optical transmission from Fig.~\ref{fig:exp_setup} is built. 

The electrical signal for the channel of interest consists of the pulse-shaped sequence for one polarization (output of the transmitter from Fig.~\ref{fig:transceiver}, which was used for simulation). It is digital-to-analog converted by a 64 GSa/s arbitrary waveform generator (AWG), and then modulates a sub-kHz linewidth fiber laser (Koheras BasiK C-15). The neighboring channels are modulated by the same modulation format as the central one by a second IQ modulator, and are fully decorrelated by more than 40 symbols using a wavelength selective switch (WSS) and delay lines. External cavity lasers (ECLs) are used as carriers on the neighboring channels. When measuring WDM performance, the fiber laser is tuned to the corresponding channel on the frequency grid (the ECLs are also tuned correspondingly). This rearrangement of the central frequencies of the lasers was performed while maintaining a constant received power for all channels in each configuration. This is achieved by controlling the input power per channel before injecting the combined signal into a recirculating loop, used for emulation of long distances. The 5 channels are then coupled and a delay-and-add polarization emulator provides the dual-polarization signal.

The loop consists of 100 km of standard, single mode fiber (SSMF) using distributed Raman amplification (DRA) with backward pumping every 50 km. In order to compensate for the power losses of the acusto-opto modulators (AOM), used as switches, an EDFA is inserted in the loop.  

The signal after transmission is detected by a four-channel 80 GSa/s coherent receiver. The local oscillator is similar to the laser at the transmitter side - a fiber laser with sub-kHz linewidth (Koheras BasiK E-15). The length of the acquired sequence corresponds to 18 blocks (similar to the simulation setup). The traces are stored, and then processed offline with the receiver from Fig.~\ref{fig:transceiver}. 

\begin{figure*}[!t]
\centering
\subfigure[Estimated received SNR]{
\label{fig:B2B:SNR}
    \includegraphics[width=2.3in]{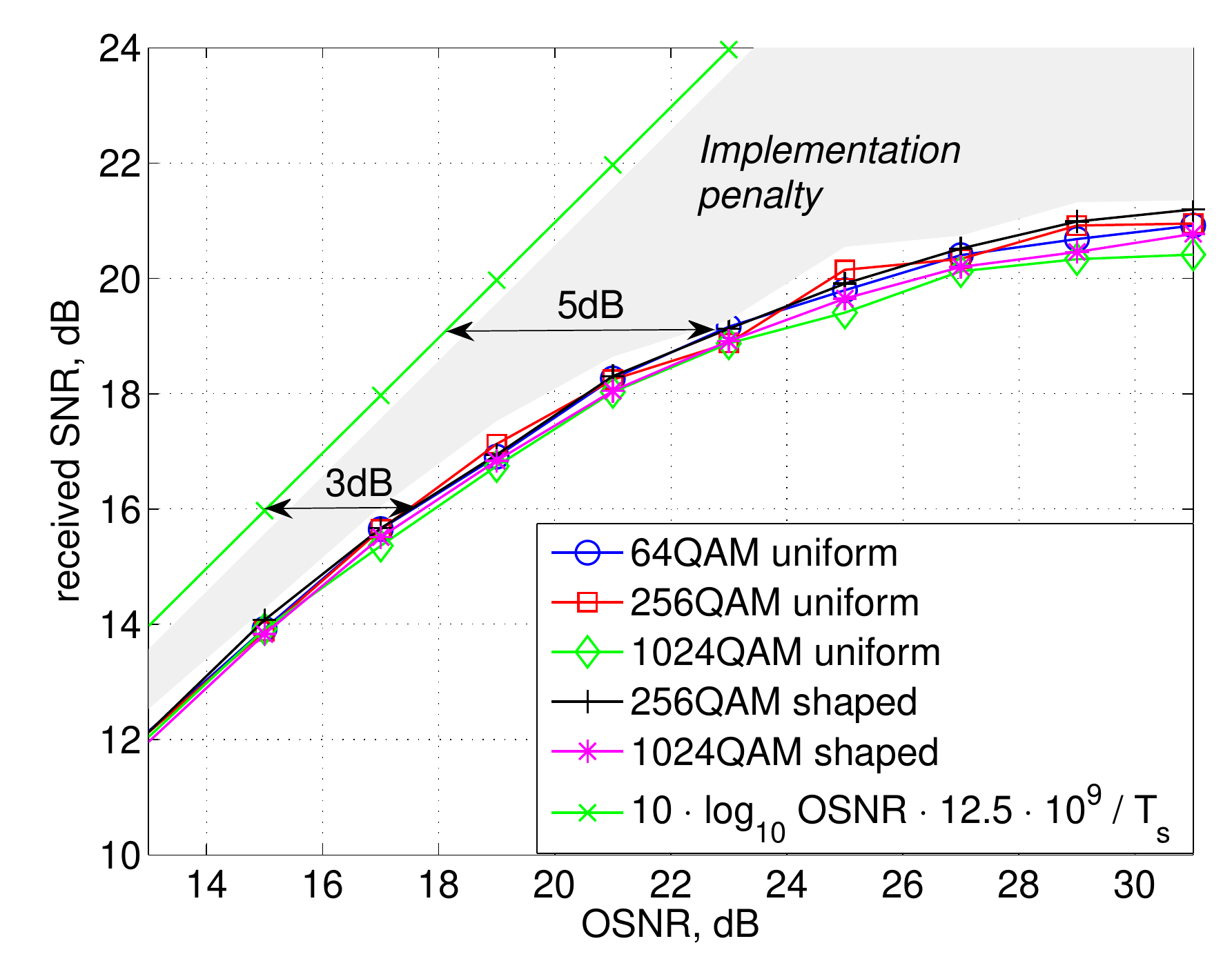}}
\subfigure[Estimated AIR]{
\label{fig:B2B:AIR}
    \includegraphics[width=2.3in]{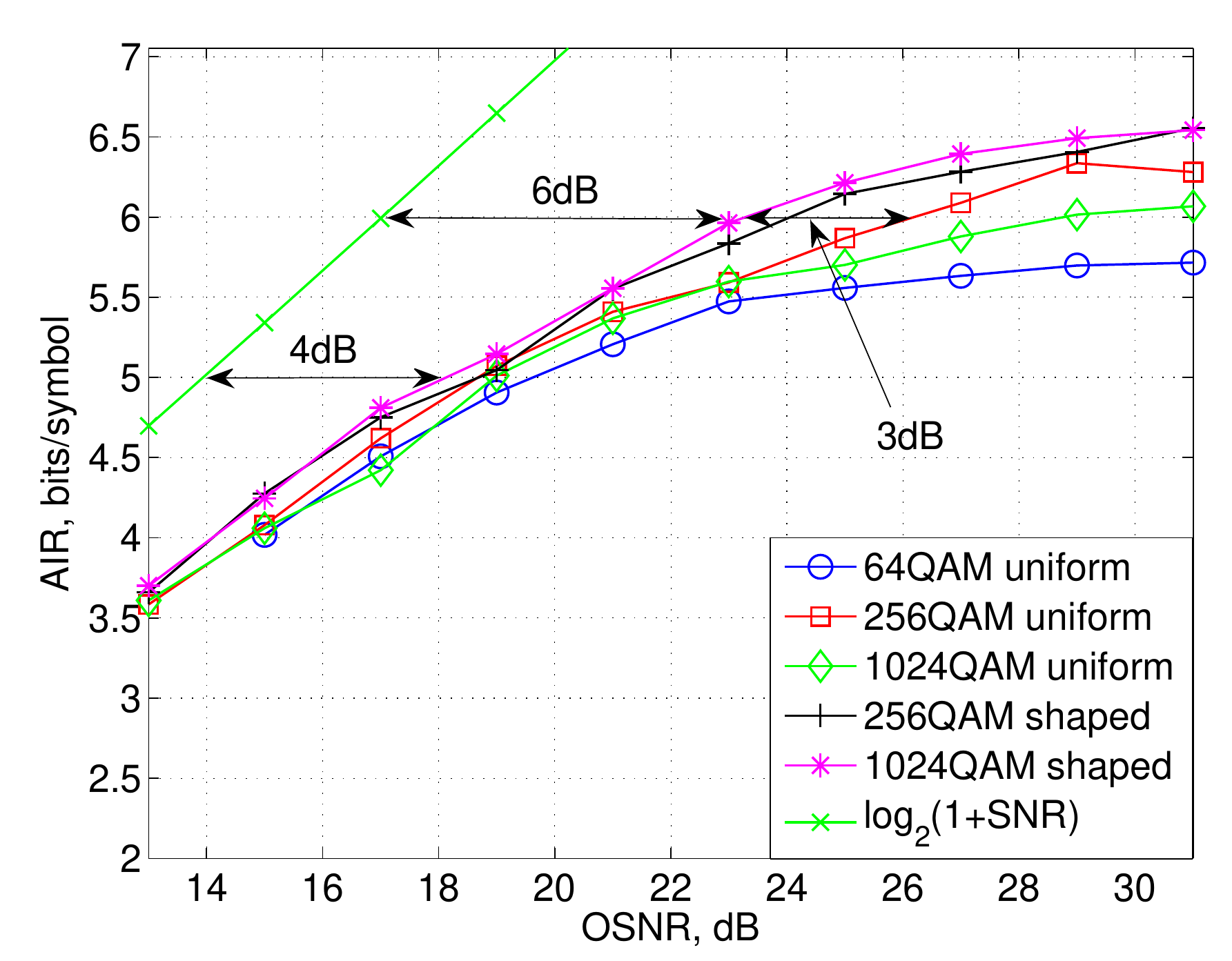}}
\subfigure[Calculated BER]{
\label{fig:B2B:BER}
    \includegraphics[width=2.3in]{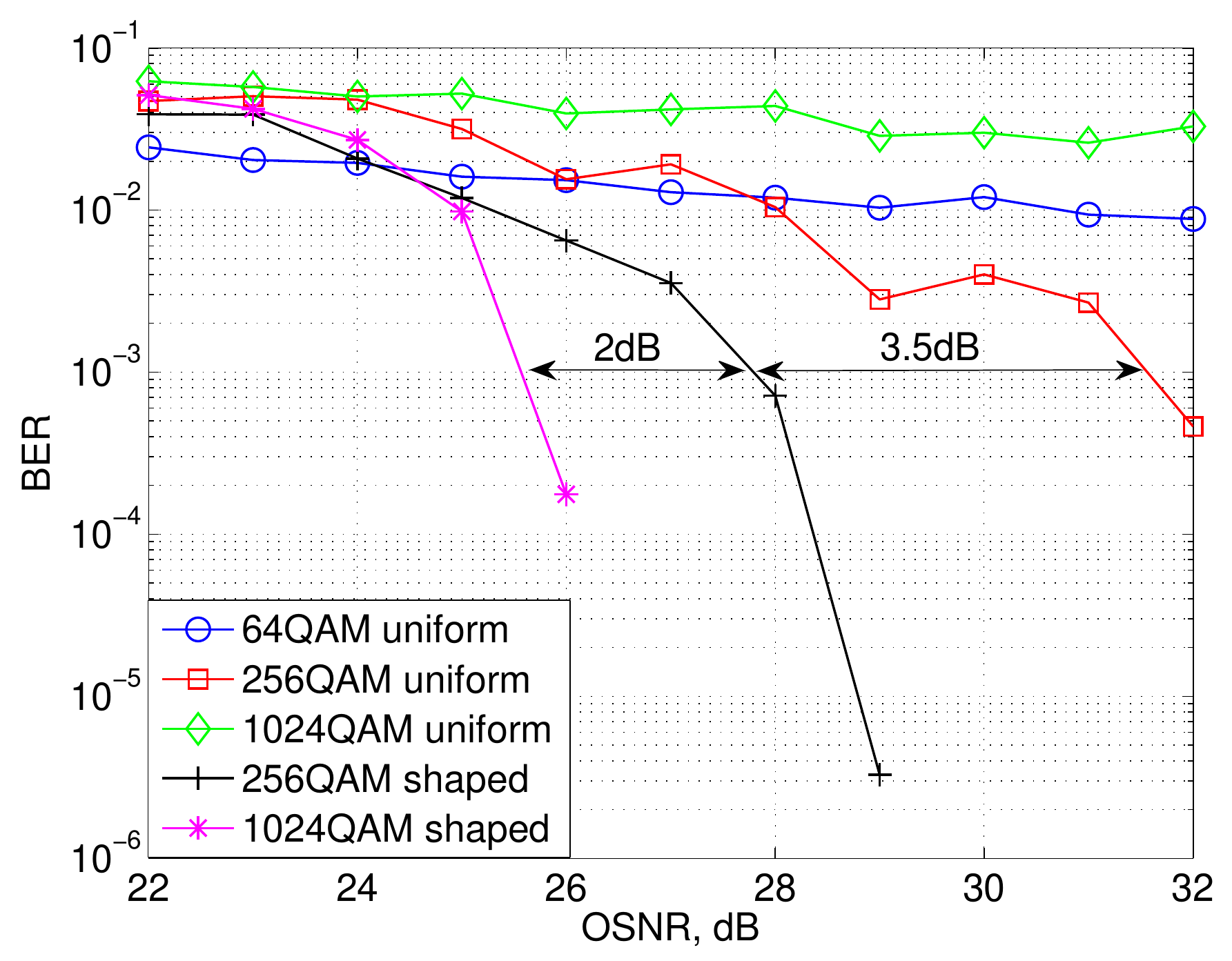}}

\caption{\textbf{Back-to-back:} Performance in optical back-to-back for the different modulation formats. The estimated SNR at high OSNR is highly dependent on the estimated noise variance, and is therefore somewhat unstable. We observe an implementation penalty of 2 dB at low OSNR and gradually increasing to around 9 dB at high OSNR.}
\label{fig:B2B}
\vspace{-.2cm}
\end{figure*}

\section{Optical back-to-back results}
Data rates of $\eta = 5, 5.5 \mbox{ and } 6$ bits/symbol are validated experimentally. Due to the limited resolution of the AWG and the sampling oscilloscope, we reduce the symbol rate to 10 GBaud and increase the roll-off factor of the pulse shape to 0.5. The pilot overhead is also increased to 2\%. The resulting data rates are given in Table~\ref{tbl:exp_rates}. The performance is shown at input data rate $\eta = 6$ bits/symbol in Fig.~\ref{fig:B2B}.

\begin{table}
\renewcommand{\arraystretch}{1.2}
\renewcommand{\tabcolsep}{15pt}
\caption{Summary of the achieved data rates experimentally}
\label{tbl:exp_rates}
\centering
\begin{tabular}{p{1cm}p{2cm}p{2cm}}
\hline
\hline
Input data rate $\eta$, [bits/symbol] & Data rate without 7\% HD FEC (BER$< 10^{-4}$), [Gbps/channel] & Data rate with 7\% HD FEC (BER$< 10^{-15}$), [Gbps/channel] \\
\hline
5 & 98 & 91.1 \\
5.5 & 107.8 & 100.2 \\
6 & 117.6 & 109.4 \\

\hline
\hline
\end{tabular}
\vspace{-.2cm}
\end{table}

The optical SNR (OSNR) is varied by an attenuator, and measured before the coherent receiver. As we see in Fig.~\ref{fig:B2B:SNR}, the received SNR is limited to about 21 dB due to the above mentioned DAC and ADC resolution. The theoretical SNR is calculated with Eq. (34) from \cite{Essiambre}, and is also given in Fig.~\ref{fig:B2B:SNR}. We observe implementation penalty of 2 dB at low OSNR and gradually increasing to around 9 dB at high OSNR (the implementation penalty here includes the sub-optimal equalization). 

In Fig.~\ref{fig:B2B:AIR}, the AIRs for the different modulation formats are shown. At $\eta = 6$ bits/symbol, the shaping gain w.r.t. 256QAM uniform is around 3 dB, and even larger w.r.t. 1024QAM uniform. The maximum received SNR is not enough to guarantee error-free uncoded transmission with 64QAM. This modulation format achieves a maximum of around 5.7 bits/symbol at the highest examined OSNR. The shaped systems proposed here operate at around $\approx$6 dB and $\approx$4 dB gap to the Shannon limit at $\eta = 6$ bits/symbol and $\eta = 5$ bits/symbol, respectively. The implementation penalty at the respective OSNR values ($\approx$23 dB and $\approx$18 dB) for these rates is $\approx$5 dB and $\approx$3 dB, which leaves $\approx$1 dB of additional penalty, which can be attributed to residual phase noise and/or sub-optimal parameter estimation (such as the mean and variance of the likelihoods). 

Finally, in Fig.~\ref{fig:B2B:BER}, the BER is shown. We see that 64QAM and 1024QAM uniform distributions are unable to obtain error-free transmission at the studied OSNR range. At BER$\approx 10^{-3}$, the 256QAM from Fig.~\ref{fig:256QAM_PMF} achieves around 3.5 dB shaping gain, which is increased to more than 5 dB with the 1024QAM from Fig.~\ref{fig:1024QAM_PMF}.

\section{WDM optical transmission results}
\subsection{Central channel}
In the last part of this paper we show the results for the recirculating loop transmission system from Fig.~\ref{fig:exp_setup}. The parameters for this case are given in Table~\ref{tbl:exp_params}. 
\begin{table}
\renewcommand{\arraystretch}{1.2}
\renewcommand{\tabcolsep}{15pt}
\caption{System parameters for WDM transmission}
\label{tbl:exp_params}
\centering
\begin{tabular}{rl}
\hline
\hline
Symbol rate & 10 GBaud \\
Pulse shape & Square-root raised cosine \\
Roll-off factor & 0.5 \\
Channel spacing & 25 GHz \\
Number of channels & 5 \\
Input data rates & 5, 5.5 and 6 bits/symbol \\
Pilot overhead & 0.02 \\

\hline
\hline
\end{tabular}
\end{table}
\begin{figure*}[!t]
\centering
\includegraphics[width=6.0in]{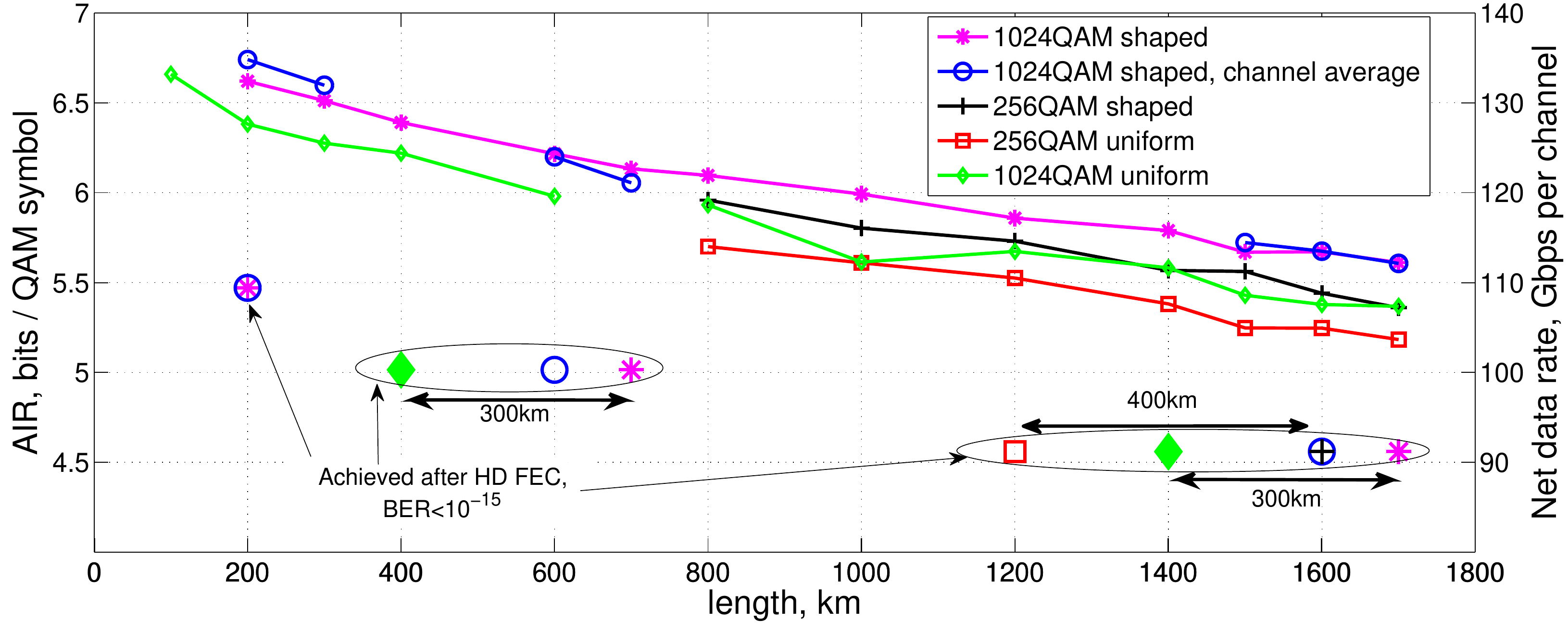}
\vspace{-.2cm}
\caption{\textbf{WDM transmission:} Maximum AIRs for the studied systems, together with their respective error-free operating points after 7\% HD FEC. The average AIR of all 5 channels of the 1024QAM shaped is also given (blue circles), together with the maximum distances, where ALL channels are error free.}
\label{fig:exp_rates}
\vspace{-.2cm}
\end{figure*}
As we saw in the previous sections, 64QAM is unreliable at data rates above 5 bits/symbol. It is therefore not included in the experimental verification. The AIRs at the optimal input power are given as a function of the studied distances in Fig.~\ref{fig:exp_rates}, together with the maximum achievable error-free distances for the different modulation formats \textit{for the central channel}. The corresponding achieved net data rates are given in Table~\ref{tbl:exp_rates}. The distances between 800 km and 1700 km and input data rate $\eta=5$ bits/symbol are studied for all modulation formats. The rates $\eta=5.5 \mbox{ and } \eta=6$ are only studied for 1024QAM shaped and uniform. Both 1024QAM shaped and 256QAM shaped provide a steady increase in the AIR of around 0.2 bits/symbol for all studied distances w.r.t. 1024QAM uniform and 256QAM uniform, respectively. We see that the error-free transmission distance can be increased by around 300 km for 1024QAM, and 400 km for 256QAM at $\eta = 5$. As we saw in Fig.~\ref{fig:B2B:BER}, the highest rate was only achieved by the shaped systems in back-to-back, and is achieved for up to 200 km with 1024QAM shaped. 

\subsection{WDM measurements for all channels}

WDM measurements are performed for the 1024QAM shaped system. The average AIR for the 5 channels in that case is also shown in Fig.~\ref{fig:exp_rates}. We see that the performance is similar for the other channels. We note again that due to lack of multiple fiber lasers, WDM measurements are performed by tuning the LO and the transmitter laser to the desired channel on the frequency grid. While the input power of each channel is optimized, so that all channels have the same received power in all such configurations, slight variations in the performance do exist for the different configurations. 

\begin{figure}[!t]
\centering
\includegraphics[width=3.2in]{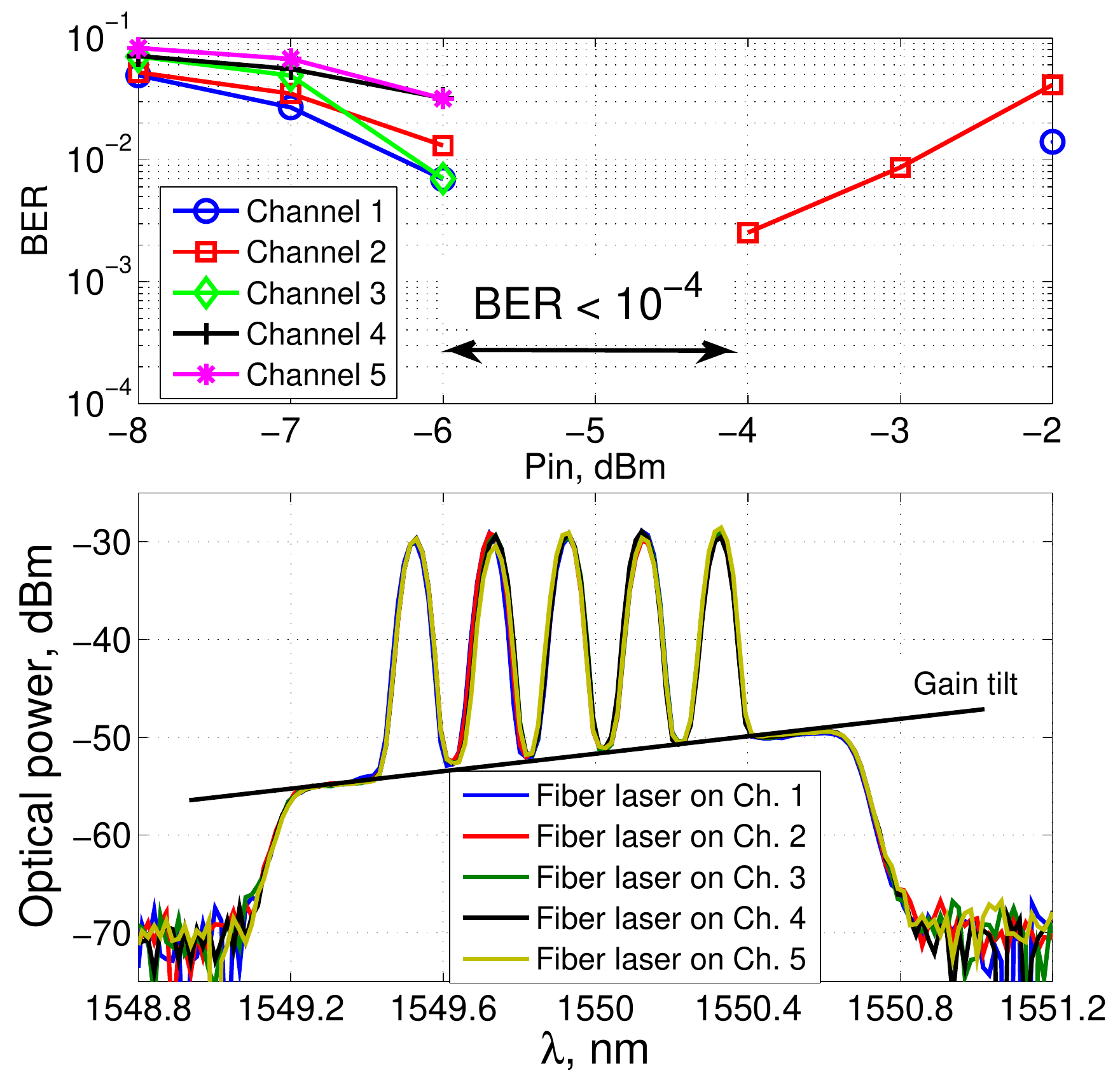}
\vspace{-.4cm}
\caption{\textbf{WDM transmission:} Measurements after 1600 km of transmission \textbf{Top:} BER as function of combined input power. \textbf{Bottom:} received spectra in each configuration of the fiber laser and the ECLs. We see that the spectra lie on top of each other, which allows us to argue that there is a fair comparison between the performance of the different channels.}
\label{fig:exp_WDM}
\vspace{-.4cm}
\end{figure}

In Fig.~\ref{fig:exp_WDM}, the received power spectra are shown for each configuration of the fiber laser and the ECLs, together with the BER on each channel for the respective configuration as a function of the \textit{combined} input power. The spectra are monitored in each case in order to ensure that the performance of the 5 channels is fairly compared. Due to the non-flat gain of the Raman pump and the non-flat ASE noise floor, the received OSNR for the different channels varies. This is because the input power is optimized for equal received power on the different channels, rather than equal OSNR. We see that due to the above mentioned gain tilt, Channel 1 (numbered from shortest to longest wavelength) has the highest OSNR, and also the best BER performance in the linear region of transmission. However, all 5 channels achieve a zero BER after turbo decoding around a combined input power of $-5$dBm. 

In general, the distances where all 5 channels achieved error free performance are somewhat shorter, than for the central channel. For example, at $\eta=5$ and $5.5$ the maximum distances are reduced to 1600 km and 600 km, respectively. For $\eta=6$ all 5 channels are error-free at 200 km. On the other hand, as we saw in Fig.~\ref{fig:exp_rates}, the channel average AIR was similar to that of the central channel, which means that some of the channels potentially achieve longer distance than the central one. Since this is a result of the unequal OSNR at the receiver, we expect similar trend for the uniform systems, and therefore similar shaping gain for all channels.

\subsection{Non-iterative demapping}
\label{sec:iterative}

Finally, we study the performance with iterative and non-iterative demapping. As can be seen from Fig.~\ref{fig:exp_rates}, the 1024QAM has highest AIR for all distances. However, due to the higher number $m$ of \textit{coded} bits per symbol, the loss from going from symbol-wise AIR to bit-wise AIR is larger. The bit-wise AIR is studied with an EXIT chart \cite{tenBrink}, which shows the extrinsic MI output of the demapper (input of the decoder, respectively) as a function of the extrinsic MI at the input of the demapper (output of the decoder, respectively). With iterative processing, the extrinsic information is exchanged between the demapper and decoder, progressing according to the chart. If the EXIT functions of the demapper and decoder do not cross, i.e. they form a tunnel in which the extrinsic information evolves with each iteration and eventually reaching '1', all bits are detected without errors. If a crossing point exists before the extrinsic information has evolved to '1', the decoder will fail and subsequent iterations will not improve the performance. 

\begin{figure}[!t]
\centering
\includegraphics[width=3.4in]{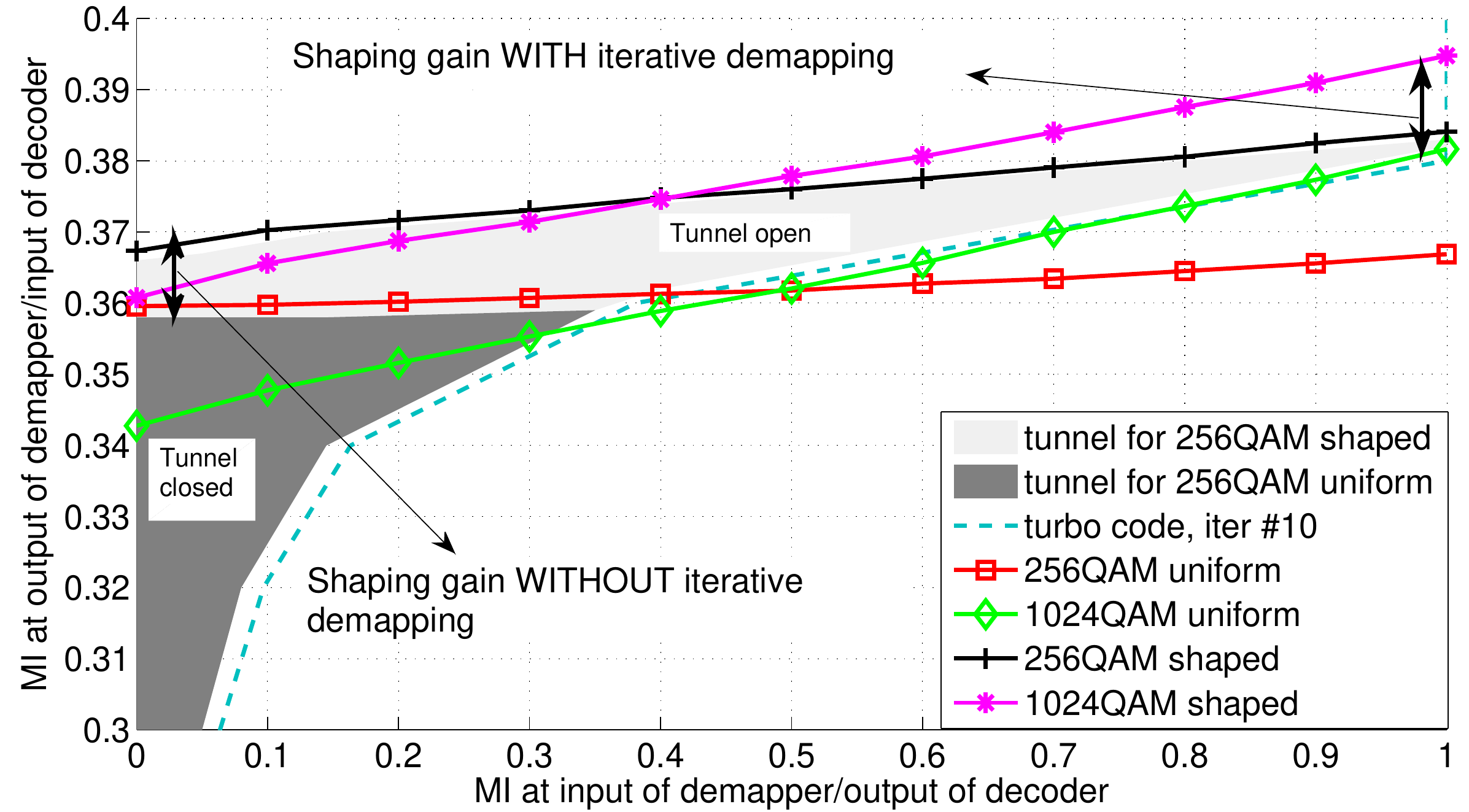}
\caption{\textbf{WDM transmission:} EXIT charts for the studied modulation formats. The shaped systems form an open tunnel with the turbo code and thus can achieve error-free performance. The smaller modulation formats have higher extrinsic information with no a-priori information and thus have better performance than the 1024QAM when non-iterative demapping is performed.}
\label{fig:EXITs_compare}
\vspace{-.7cm}
\end{figure}

An example of the \textit{experimental} EXIT chart for the proposed system is given in Fig.~\ref{fig:EXITs_compare}. The experimental EXIT functions of the demappers are obtained by calculating the log-likelihoods of the symbols after phase noise tracking, then generating a-priori information via the Gaussian model\cite{tenBrink}. The input power for this chart is $-3$ dBm and the distance is 1400 km. Since the FEC rate is different for the different formats at fixed net data rate, the normalization of the bitwise MI is performed w.r.t. a fixed gross data rate, where the punctured bits (if any) provide no MI, allowing for fair comparison assuming the minimum FEC rate of $R=1/3$. As expected, the shaped systems provide a larger area below their EXIT functions, which indicates that higher data rate is potentially possible with proper selection of FEC code. We see that in contrast to their symbol-wise MI (see AIR in Fig. \ref{fig:exp_rates}), the bit-wise MI of the 1024QAM shaped is lower than that of 256QAM shaped when no a-priori information is available to the demapper. This case corresponds to non-iterative demapping, where 256QAM can therefore be expected to outperform 1024QAM. However, when the decoder feeds extrinsic information back to the demapper, the steeper inclination of the 1024QAM results in higher gains from iterative processing and better overall performance. Both shaped systems cross the EXIT function of the turbo decoder after they have achieved a MI of '1', which implies correct decoding. The EXIT function of the 256QAM uniform starts low and is almost flat, which results in early crossing of the EXIT function of the turbo code. This would result in erroneous decoding, which cannot be improved with iterations. 

\begin{figure}[!t]
\centering
\includegraphics[width=3.3in]{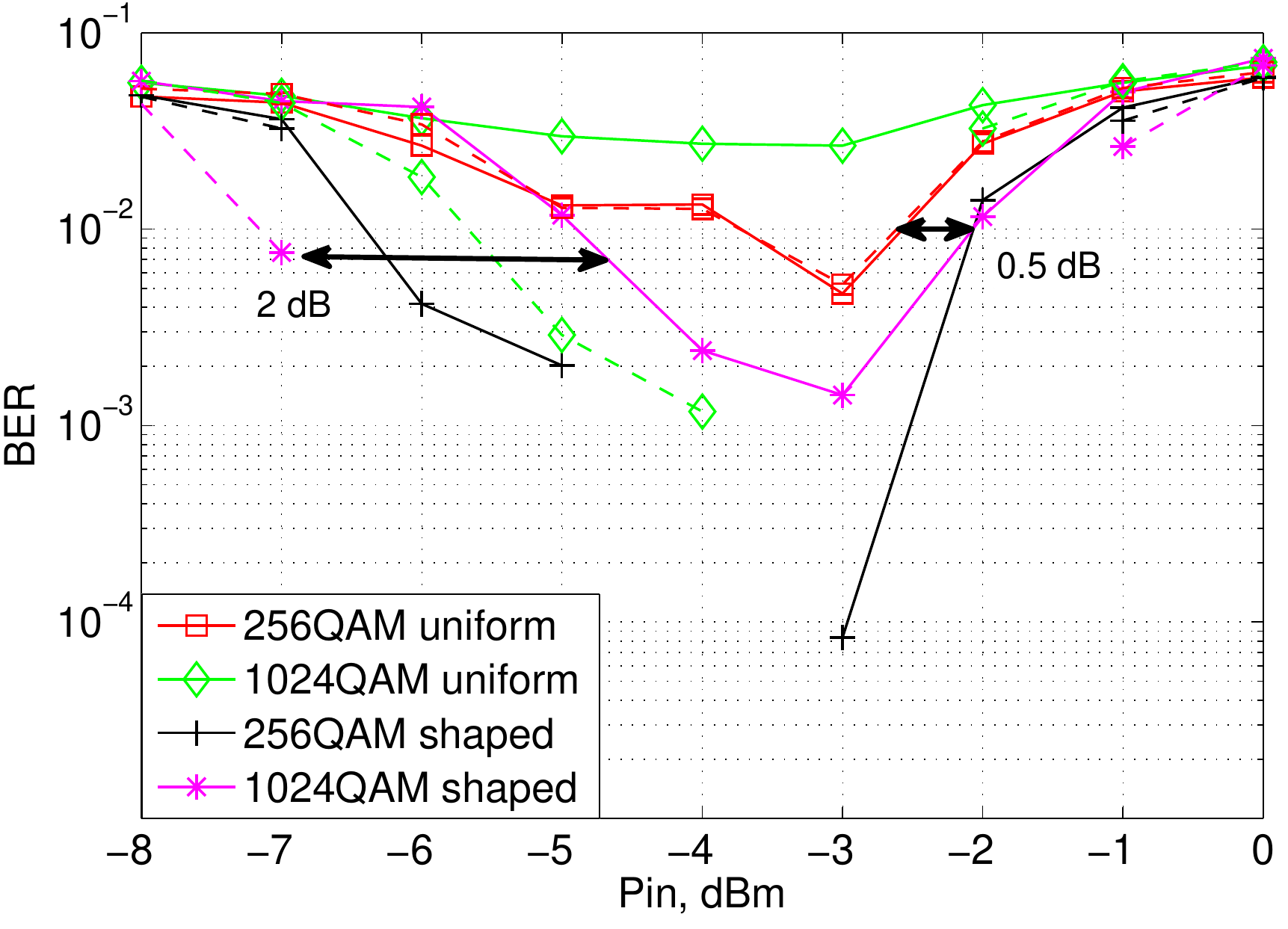}
\vspace{-.5cm}
\caption{\textbf{WDM transmission:} Performance with iterative (dashed lines) and non-iterative (solid lines) demapping for the central channel after 1400 km. The smaller modulation formats generally perform better in the latter case (explanation in the text).}
\label{fig:exp_iterative_1400}
\vspace{-.6cm}
\end{figure}

In Fig.~\ref{fig:exp_iterative_1400}, the BER performance in the case of iterative (dashed lines) and non-iterative (solid lines) demapping and decoding are shown after 1400 km. As predicted above, the 256QAM performs better than 1024QAM without iterations for both shaped and non-shaped systems. Shaping gain is however still achieved over the 256QAM uniform, in this case - around 1.5 dB and 0.5 dB in the linear and non-linear region, respectively, at BER$=10^{-2}$. This translates to 200 km of increased maximum distance (for the measurements we have). When iterations are performed, the performance of 1024QAM shaped is improved by more than 2 dB in the linear and around 0.5 dB in the non-linear region, making it superior to the other modulation formats. The same principle follows for the uniform distributions - the 256QAM hardly gains from iterative demapping, whereas 1024QAM uniform is able to achieve an error-free performance at this distance.

\section{Discussion and future work}
The main drawback of the proposed method is the integration with turbo codes, which suffer from high latency. Even though multi-Gbps hardware implementations exist for turbo codes \cite{Shrestha}, a large degree of parallelization will be needed in order to achieve the hundreds of Gbps per channel given in Fig.~\ref{fig:sim_rates} and Fig.~\ref{fig:exp_rates}. Another solution would be to decrease the symbol rate per channel, and thereby the necessary serial bit rate. We note that the optimized PMF for significantly different symbol rates may be different, and the optimization procedure from Section \ref{sec:optimization} should be used.

In this work, the codeword length was constant for different rates for easier software implementation. However, fixed clock speed is preferable for hardware implementation, which can be realized by fixing the information sequence length and varying the codeword length instead, resulting in variable decoder throughput at the expense of slight dependence of the decoding performance on the data rate. 

As discussed in Section \ref{sec:iterative}, non-iterative demapping is possible at the cost of degraded performance. It would be of interest to perform optimization of the scheduling procedure of the iterations between the decoder and demapper and the internal decoder iterations, in order to achieve a better trade-off between performance and complexity/latency. We note that turbo codes of lower complexity, e.g. with shorter memory of the convolutional encoders, can also be used without sacrificing the shaping gain, but slightly degrading the overall performance.

An interesting area for future work is to integrate the mapping functions, designed here, with another type of near-capacity achieving channel code, e.g. LDPC convolutional codes (LDPC-CC) (also known as spatially coupled LDPC codes). Pipe-lining structures exist for LDPC-CC \cite{Arikan}, allowing for decoding of very fast serial data streams with reasonable latency. The main challenge in this integration is that bit ambiguities may be considered as puncturing, which effectively inserts cycles in the LDPC code graph, thereby hindering the performance of belief propagation decoding. The puncturing patterns would therefore have to be very carefully designed. Alternatively, a combination of the PMFs, designed in this paper with the rate-adaptive system from \cite{Buchali, Buchali-2, Bocherer-2} would be of interest for potentially improving the AIRs while avoiding puncturing.

We conclude this section with a note on the complexity of the many-to-one demappers w.r.t. standard QAM demappers. The difference here is that generally more bits per symbol $m$ need to be demapped. The complexity of a QAM demapper scales linearly with $m$ as ${\cal O}(m)$ \cite{Wang}, and is furthermore negligible compared to the other parts of the receiver, making the added complexity from many-to-one mapping and demapping negligible. The proposed system is also particularly easy to integrate with systems, already using turbo coded modulation. 

\section{Conclusion}
In this paper, a method for probabilistic shaping for wavelength division multiplexed (WDM) fiber optic communications was proposed. The method relies on many-to-one mapping functions to achieve an optimized probability mass function (PMF) of the channel input. Rate adaptation was performed by punctured turbo codes, which allows for shaping gains to be achieved on a wide range of distances and/or data rates without changing the PMF. 

The method was experimentally demonstrated in a 5 channel dual polarization WDM system at 10 GBaud per channel. For data rates between 5 and 6 bits/symbol, the proposed mappings provided an increase of around 0.2 bits/symbol, which translates to around 4 Gbps/channel. Alternatively, the maximum achieved distance at fixed rate was increased by around 300 km for all studied rates and distances, which is between 75\% and 20\% for short ($<$500 km) and medium-to-long ($>$1000 km) distances, respectively.  

\section{Acknowledgments}
\small{This work was supported by the DNRF Research Centre of Excellence, SPOC, ref. DNRF123. NKT Photonics A/S is acknowledged for providing the narrow linewidth fiber lasers used in the experiment.} 

\bibliographystyle{IEEEtran}
\bibliography{refs}

\begin{thebibliography}{10}
\providecommand{\url}[1]{#1}
\csname url@samestyle\endcsname
\providecommand{\newblock}{\relax}
\providecommand{\bibinfo}[2]{#2}
\providecommand{\BIBentrySTDinterwordspacing}{\spaceskip=0pt\relax}
\providecommand{\BIBentryALTinterwordstretchfactor}{4}
\providecommand{\BIBentryALTinterwordspacing}{\spaceskip=\fontdimen2\font plus
\BIBentryALTinterwordstretchfactor\fontdimen3\font minus
  \fontdimen4\font\relax}
\providecommand{\BIBforeignlanguage}[2]{{%
\expandafter\ifx\csname l@#1\endcsname\relax
\typeout{** WARNING: IEEEtran.bst: No hyphenation pattern has been}%
\typeout{** loaded for the language `#1'. Using the pattern for}%
\typeout{** the default language instead.}%
\else
\language=\csname l@#1\endcsname
\fi
#2}}
\providecommand{\BIBdecl}{\relax}
\BIBdecl

\bibitem{Cover}
T.~M. Cover and J.~A. Thomas, \emph{Elements of Information Theory, 2nd
  edition}.\hskip 1em plus 0.5em minus 0.4em\relax Hoboken, NJ: John Wiley \&
  Sons, Inc., 2006.

\bibitem{Essiambre}
R.-J. Essiambre, G.~Kramer, P.~J. Winzer, G.~J. Foschini, and B.~Goebel,
  ``Capacity limits of optical fiber networks,'' \emph{IEEE Journal of
  Lightwave Technology}, vol.~28, no.~4, pp. 662--701, Feb. 2010.

\bibitem{Dar-shaping}
R.~Dar, M.~Feder, A.~Mecozzi, and M.~Shtaif, ``On shaping gain in the nonlinear
  fiber-optic channel,'' in \emph{Proc. of IEEE International Symposium on
  Information Theory}, July 2014, pp. 2794--2798.

\bibitem{Smith}
B.~P. Smith and F.~R. Kschischang, ``A pragmatic coded modulation scheme for
  high-spectral-efficiency fiber-optic communications,'' \emph{IEEE Journal of
  Lightwave Technology}, vol.~30, no.~13, pp. 2047--2053, July 2012.

\bibitem{Fehen}
T.~Fehenberger, G.~B\"ocherer, A.~Alvarado, and N.~Hanik, ``{LDPC} coded
  modulation with probabilistic shaping for optical fiber systems,'' in
  \emph{Proc. of Optical Fiber Communication Conference (OFC)}, Mar. 2015, p.
  Th2A.23.

\bibitem{Agrell}
L.~Beygi, E.~Agrell, J.~M. Kahn, and M.~Karlsson, ``Rate-adaptive coded
  modulation for fiber-optic communications,'' \emph{IEEE Journal of Lightwave
  Technology}, vol.~32, no.~2, pp. 333--343, Jan. 2014.

\bibitem{Buchali}
F.~Buchali, G.~B\"ocherer, W.~Idler, L.~Schmalen, P.~Schulte, and F.~Steiner,
  ``Experimental demonstration of capacity increase and rate-adaptation by
  probabilistically shaped 64-{QAM},'' in \emph{Proc. of European Conference on
  Optical Communications (ECOC)}, Oct. 2015, p. PDP.3.4.

\bibitem{Buchali-2}
F.~Buchali, F.~Steiner, G.~B\"ocherer, L.~Schmalen, P.~Schulte, and W.~Idler,
  ``Rate adaptation and reach increase by probabilistically shaped 64-{QAM}: An
  experimental demonstration,'' \emph{IEEE Journal of Lightwave Technology},
  vol.~34, no.~7, pp. 1599--1609, Apr. 2016.

\bibitem{Lotz}
T.~H. Lotz, X.~Liu, S.~Chandrasekhar, P.~J. Winzer, H.~Haunstein, S.~Randel,
  S.~Cortesilli, B.~Zhu, and D.~W. Peckham, ``Coded {PDM-OFDM} transmission
  with shaped 256-iterative-polar-modulation achieving 11.15-b/s/hz
  intrachannel spectral efficiency and 800-km reach,'' \emph{IEEE Journal of
  Lightwave Technology}, vol.~31, no.~4, pp. 528--545, Jan. 2013.

\bibitem{Estaran}
J.~Estar\'an, D.~Zibar, and I.~T. Monroy, ``Capacity-approaching superposition
  coding for optical fiber links,'' \emph{IEEE Journal of Lightwave
  Technology}, vol.~32, no.~17, pp. 2960--2972, Sep. 2014.

\bibitem{Liu}
T.~Liu and I.~B. Djordjevic, ``Optimal signal constellation design for
  ultrahigh-speed optical transport in the presence of nonlinear phase noise,''
  \emph{Optics Express}, vol.~22, no.~26, pp. 32\,188--32\,198, Dec. 2014.

\bibitem{Djordjevic}
I.~B. Djordjevic, A.~Z. Jovanovic, Z.~H. Peric, and T.~Wang, ``Multidimensional
  optical transport based on optimized vector-quantization-inspired signal
  constellation design,'' \emph{IEEE Transactions on Communications}, vol.~62,
  no.~9, pp. 3262--3273, Sep. 2014.

\bibitem{Yankov}
M.~P. Yankov, S.~Forchhammer, K.~J. Larsen, and L.~P.~B. Christensen,
  ``Rate-adaptive constellation shaping for turbo-coded {BICM},'' in
  \emph{Proc. of IEEE International Conference on Communications (ICC)}, June.
  2014, pp. 2112--2117.

\bibitem{Yankov-PTL}
M.~P. Yankov, D.~Zibar, S.~Forchhammer, K.~J. Larsen, and L.~P.~B. Christensen,
  ``Constellation shaping for fiber-optic channels with {QAM} and high spectral
  efficiency,'' \emph{IEEE Photonics Technology Letters}, vol.~26, no.~23, pp.
  2407--2410, Dec. 2014.

\bibitem{Arnold}
D.~M. Arnold, H.-A. Loeliger, P.~O. Vontobel, A.~Kav\v{c}i\'c, and W.~Zeng,
  ``Simulation-based computation of information rates for channels with
  memory,'' \emph{IEEE Transactions on Information Theory}, vol.~52, no.~8, pp.
  3498--3508, Aug. 2006.

\bibitem{Varnica}
N.~Varnica, X.~Ma, and A.~Kav\v{c}i\'c, ``Capacity of power constrained
  memoryless {AWGN} channels with fixed input constellations,'' in \emph{Proc.
  of {GLOBECOM}}, Nov. 2002, pp. 1339--1343.

\bibitem{Eriksson}
\BIBentryALTinterwordspacing
T.~A. Eriksson, T.~Fehenberger, P.~A. Andrekson, M.~Karlsson, N.~Hanik, and
  E.~Agrell, ``Impact of 4d channel distribution on the achievable rates in
  coherent optical communication experiments,'' Nov. 2015. [Online]. Available:
  \url{http://arxiv.org/abs/1512.02512}
\BIBentrySTDinterwordspacing

\bibitem{Dar_invited}
R.~Dar, M.~Feder, A.~Mecozzi, and M.~Shtaif, ``Inter-channel nonlinear
  interference noise in {WDM} systems : Modeling and mitigation,'' \emph{IEEE
  Journal of Lightwave Technology}, vol.~33, no.~5, pp. 1044--1053, Mar. 2015.

\bibitem{Carena_EGN}
A.~Carena, G.~Bosco, V.~Curri, Y.~Jiang, P.~Poggiolini, and F.~Forghieri,
  ``{EGN} model of non-linear fiber propagation,'' \emph{Optics Express},
  vol.~22, no.~13, pp. 16\,335--16\,362, June 2014.

\bibitem{Bocherer-2}
G.~B\"ocherer, F.~Steiner, and P.~Schulte, ``Bandwidth efficient and
  rate-matched low-density parity-check coded modulation,'' \emph{IEEE
  Transactions on Communications}, vol.~63, no.~12, pp. 4651--4665, Oct. 2015.

\bibitem{Raphaeli}
D.~Raphaeli and A.~Gurevitz, ``Constellation shaping for pragmatic turbo-coded
  modulation with high spectral efficiency,'' \emph{IEEE Transactions on
  Communications}, vol.~52, no.~3, pp. 341--345, Mar. 2004.

\bibitem{Bocherer}
G.~B\"ocherer, F.~Altenbach, and R.~Mathar, ``Capacity achieving modulation for
  fixed constellations with average power constraint,'' in \emph{Proc. of IEEE
  International Conference on Communications}, June. 2011, pp. 1--5.

\bibitem{Zepernick}
H.-J. Zepernick and A.~Finger, \emph{Pseudo random signal processing : theory
  and application}.\hskip 1em plus 0.5em minus 0.4em\relax The Atrium, Southern
  Gate, Chichester, West Sussex: John Wiley and sons, inc., 2005.

\bibitem{CDcomp}
S.~J. Savory, ``Digital filters for coherent optical receivers,'' \emph{Optics
  Express}, vol.~16, no.~2, pp. 804--817, Jan. 2008.

\bibitem{Yankov-PN}
M.~P. Yankov, T.~Fehenberger, L.~Barletta, and N.~Hanik, ``Low-complexity
  tracking of laser and nonlinear phase noise in {WDM} optical fiber systems,''
  \emph{IEEE Journal of Lightwave Technology}, vol.~33, no.~23, pp. 4975--4984,
  Dec. 2015.

\bibitem{tenBrink}
S.~ten Brink, ``Convergence behavior of iteratively decoded parallel
  concatenated codes,'' \emph{IEEE Transactions on Communications}, vol.~49,
  no.~10, pp. 1727--1737, Oct. 2001.

\bibitem{Shrestha}
R.~Shrestha and R.~P. Paily, ``High-throughput turbo decoder with parallel
  architecture for {LTE} wireless communication standards,'' \emph{IEEE
  Transactions on Circuits and Systems - I}, vol.~61, no.~9, pp. 2699--2710,
  Sep. 2014.

\bibitem{Arikan}
E.~Arikan, N.~Hassan, M.~Lentmaier, G.~Montorsi, and J.~Sayir, ``Challenges and
  some new directions in channel coding,'' \emph{Journal of Communications and
  Networks}, vol.~17, no.~4, pp. 328--338, Aug. 2015.

\bibitem{Wang}
Q.~Wang, Q.~Xie, Z.~Wang, C.~Sheng, and L.~Hanzo, ``A universal low-complexity
  symbol-to-bit soft demapper,'' \emph{IEEE Transactions on Vehicular
  Technology}, vol.~63, no.~1, pp. 119--130, Jan. 2014.

\end{thebibliography}

\end{document}